\documentclass[a4paper,11pt,onecolumn]{article}%
\usepackage{amsmath}
\usepackage{amsfonts}
\usepackage{amssymb}
\usepackage{graphicx}
\usepackage{bm}
\usepackage{geometry}%
\providecommand{\U}[1]{\protect\rule{.1in}{.1in}}
\providecommand{\U}[1]{\protect\rule{.1in}{.1in}}
\providecommand{\U}[1]{\protect\rule{.1in}{.1in}}
\providecommand{\U}[1]{\protect\rule{.1in}{.1in}}
\providecommand{\U}[1]{\protect\rule{.1in}{.1in}}
\providecommand{\U}[1]{\protect\rule{.1in}{.1in}}
\providecommand{\U}[1]{\protect\rule{.1in}{.1in}}
\providecommand{\U}[1]{\protect\rule{.1in}{.1in}}
\providecommand{\U}[1]{\protect\rule{.1in}{.1in}}
\providecommand{\U}[1]{\protect\rule{.1in}{.1in}}
\providecommand{\U}[1]{\protect\rule{.1in}{.1in}}
\providecommand{\U}[1]{\protect\rule{.1in}{.1in}}
\providecommand{\U}[1]{\protect\rule{.1in}{.1in}}
\providecommand{\U}[1]{\protect\rule{.1in}{.1in}}
\providecommand{\U}[1]{\protect\rule{.1in}{.1in}}
\providecommand{\U}[1]{\protect\rule{.1in}{.1in}}
\providecommand{\U}[1]{\protect\rule{.1in}{.1in}}
\begin{document}

\title{Surrealistic Bohmian trajectories do not occur with macroscopic pointers}
\author{G. Tastevin\footnote{tastevin@lkb.ens.fr} and F. Lalo\"{e}\footnote{laloe@lkb.ens.fr}\\Laboratoire Kastler Brossel, ENS-Universit\'e PSL,\\ CNRS, Sorbonne Universit\'e, Coll\`ege de France,\\24 rue Lhomond 75005\ Paris, France}
\date{\today}
\maketitle

\begin{abstract}
We discuss whether position measurements in quantum mechanics can be
contradictory with Bohmian trajectories, leading to what has been called
\textquotedblleft surrealistic trajectories\textquotedblright\ in the
literature. Previous work has considered that
a single Bohmian position can be ascribed to the pointer. Nevertheless, a
correct treatment of a macroscopic pointer requires that many particle
positions should be included in the dynamics of the system, and that
statistical averages should be made over their random initial values. Using
numerical as well as analytical calculations, we show that these surrealistic
trajectories exist only if the pointer contains a small number of particles;
they completely disappear with macroscopic pointers.

With microscopic pointers, non-local effects of quantum entanglement can indeed take place and
introduce unexpected trajectories, as in Bell experiments; moreover, the initial values of the
Bohmian positions associated with the measurement apparatus may influence the
trajectory of the test particle, and determine the result of measurement.
Nevertheless, a detailed observation of the trajectories of the particles of
the pointer reveals the nature of the trajectory of the test
particle; nothing looks surrealistic if all trajectories are properly interpreted.

\end{abstract}

\begin{center}
\ **********
\end{center}


De Broglie and Bohm (dBB) have introduced an interpretation of quantum
mechanics where material particles are described, not only by wave functions
as in standard quantum mechanics, but also by point positions that are guided
by the wave function \cite{De-Broglie-1927, Bohm-1952}. Trajectories in the
configuration space can then be associated with the time evolution of any
system made of massive particles.\ These trajectories can be projected into
ordinary 3D space, which provides the trajectory of each constituent particle.
Such projections sometimes exhibit unexpected properties. They are interesting, since their study may reveal unexpected quantum
properties that could otherwise remain unnoticed inside the standard
equations. General reviews of Bohmian mechanics and trajectories in various situations can
be found for instance in \cite{Holland-1993}, \cite{Oriols-Mompart} or \cite{Bricmont-2016}.

In the context of standard theory, the study of these trajectories
leads to nothing but a visualization of the motion of the usual probability
current.  Because the dBB language is convenient, in this article we will speak of positions and trajectories of the
particles, instead of streamlines of the probability. Nevertheless, the reader
who is allergic to the very idea of particle positions in quantum mechanics
can easily translate every statement in terms of the trajectories of the
elements of the probability fluid.\ Our purpose here is not to plead in favor
of one interpretation or another, but just to provide a more precise analysis
of the dBB trajectories in the presence of quantum non-local effects. In particular,
we will not study the trajectories defined within the consistent history
interpretation \cite{Griffiths-1999}, which are different.

In 1992, Englert, Scully, S\"{u}ssmann and Walther proposed an interesting
thought experiment \cite{Englert-1992} where a test particle crosses a two
slit interferometer, while another quantum system (the \textquotedblleft
pointer\textquotedblright) plays the role of a \textquotedblleft Welcher
Weg\textquotedblright\ (which way) detector, indicating through which slit the
test particle went. These authors argue that, if one observes the position of
the pointer after the test particle has crossed the interference region, in
some cases the pointer seems to indicate that the particle crossed one slit,
while a reconstruction of the past Bohmian trajectory of the particles shows
that it went through the other.\ The reason behind this unexpected conclusion
is that, when two wave packets of the test particle cross in the interference
region, the Bohmian position of the particle may \textquotedblleft
jump\textquotedblright\ from one wave packet to the other, leading to a curved
Bohmian trajectory without any force from outside.\ The authors of
\cite{Englert-1992} consider that, even under these conditions, the indication of the pointer still
provides a correct measurement of which slit was really crossed by the test particle; since some Bohmian trajectories nevertheless cross the other slit, they express strong
doubts about the real physical interest of these trajectories, and call them
\textquotedblleft surrealistic\textquotedblright.

Several authors have discussed the question and reached various
conclusions.\ A first reaction by D\"{u}rr et al. \cite{Durr-et-al-1993} in
1993 was very general: these authors pointed out that this property of Bohmian
trajectories is no more paradoxical than the orthodox point of view where the
particles goes through two slits at the same time; in their opinion, the
qualification of \textquotedblleft surrealistic" could arise only from a naive
version of operationalism. Also in 1993, Dewdney, Hardy and Squires published
an article \cite{Dewdney-et-al-1993} where a Mach-Zhender interferometer is
studied; they argue that this simpler version of the thought experiment
reveals nothing surrealistic about Bohmian trajectories, but only illustrates
the well-known non-local influence of the quantum potential introduced by Bohm
\cite{Bohm-1952}. Three years later, Aharonov and Vaidman discussed the
relation between position measurements and Bohmian positions in general, but
also in the context of weak and protective measurements
\cite{Aharonov-Vaidman-1996}. In 1998, Scully stated again that, in his
opinion, Bohmian trajectories do not always provide a trustworthy physical
picture of particle motion \cite{Scully-1998}; he nevertheless concludes that
he agrees Bohmian mechanics offers an interesting line of
thought. In 1999, Aharonov,\ Englert and Scully studied the relation between
\textquotedblleft protective measurements\textquotedblright\ and Bohmian
trajectories, and concluded that the results challenge any realistic
interpretation of the trajectories \cite{Aharonov-Englert-Scully-1999}.\ But,
in 2006, Hiley expressed again the opinion that surreal trajectories occur
only from an incorrect use of the Bohm approach \cite{Hiley-2006}. In 2007,
Wiseman pointed out that the measurement of weak values can lead to an
experimental determination of Bohmian trajectories \cite{Wiseman--2007}; see
also Ref.\ \cite{Durr-Goldstein-Zanghi-2009}.\ Bohmian trajectories can also be
reconstructed with a generalization of the technique of quantum state
reconstruction \cite{Schleich-Freyberger-Zubairi-2013}. Gisin
\cite{Gisin-2015} has pointed out that surrealistic trajectories occur only if
the pointer of the measurement apparatus is \textquotedblleft
slow\textquotedblright, i.e. if it indicates which slit the particle crossed
with a delay, only after the particle has left the interference
region.\ We conclude this brief review by noting that related experiments have
been performed.\ In 2011, Kocsis et al.\ have observed the average
trajectories of single photons in a two slit interferometer
\cite{Kocsis-et-al-2011}.\ Of course, photons are relativistic particles,
which stricto sensu have no quantum position operator, and therefore neither a
Bohmian position nor a trajectory.\ Nevertheless, Braveman and Simon
\cite{Braverman-Simon-2013} have pointed out that, in the paraxial
approximation, the propagation of an electromagnetic field obeys an equation
that is identical to the Schr\"{o}dinger equation for a 2D massive particle;
the same authors have proposed to observe the non-locality of trajectories
with entangled photons.\ In 2016, Mahler et al.\ have indeed observed non-local
and surreal Bohmian trajectories, and conclude that the trajectories seem
surreal only if one ignores their manifest non-locality
\cite{Mahler-et-al-2016}.

A common feature of these publications is that their theoretical analysis
assigns a single Bohmian position to the pointer of the measurement apparatus;
the corresponding object may for instance be the center of mass of the
pointer, and therefore have a macroscopic mass. Nevertheless, in a real
measurement apparatus, the pointer necessarily contains many particles, which
are described by a large number of degrees of freedom.\ It then becomes
necessary to assign a large number of Bohmian positions to the pointer; the purpose of this article is to discuss their role.\ As we will see, their presence radically changes
the situation, since the fraction of \textquotedblleft
surrealistic\textquotedblright\ trajectories decreases when the number of
positions increases.\ In other words, we will check that non-local
effects vanish in the macroscopic limit for the pointer, as one could expect.
Moreover, even with microscopic pointers, if quantum effects are properly taken
into account, no actual contradiction appears between the trajectories and the
indications given by the pointer. Actually, the pointer positions can be used to obtain correct
information, not only on the slit crossed by the test particle, but also
quantum non-local effects taking place in the interference region.

In \S ~\ref{one-particle-pointer}, we introduce the model used to make the
numerical computations in this article.\ In the original scheme of
Ref.~\cite{Englert-1992}, the Welcher Weg quantum system was a micromaser
enclosed in a cavity, possibly complemented by a second massive particle taking
a trajectory that depends on the state of radiation inside the
cavity.\ In order to avoid the introduction of Bohmian variables associated
with the electromagnetic field, in this article we will assume that the
Welcher Weg apparatus contains only (one or several) massive particles having
Bohmian positions. In \S ~\ref{fast-and-slow-pointers}, we assume that the
pointer is made of a single particle and we discuss the characteristics of the
trajectories in different situations: pointers providing Welcher Weg
information almost immediately, or only delayed information, or intermediate cases.\ In
\S ~\ref{pointer-several-particles}, we assume that one or two pointers contain several particles,
each of them associated with its Bohmian
position.\ Numerical calculations then show that the trajectories strongly
depend on the number of these positions; quantum non-local effects may still
take place only if the number of Bohmian positions is not too high. In
\S ~\ref{macroscopic-pointer}, the pointer is assumed to be macroscopic; it
contains an enormous number of particles (some fraction of the Avogadro
number).\ An analytic argument shows that non-local effects then disappear:
surrealistic Bohmian trajectories no longer exist when the pointer is
macroscopic (this analytic argument is expanded in an Appendix, where we show
that this disappearance is a consequence of the multiplication of the effective
velocity of the pointer by a factor  $\sqrt{N}$, where $N$ is the number of
particles $N$ contained in the pointer). Finally, in \S ~\ref{discussion}, we
conclude that adequate observations of the pointer trajectories allows one to
understand the detailed characteristics of the test particle trajectory in all cases: as
soon as the non-local dynamics of the coupled quantum systems is well understood, the
trajectories cease to look surrealistic. A preliminary account of this work
can be found in~\cite{FL-CNVLMQ}.

\section{Pointer with one degree of freedom, numerical model}
\label{one-particle-pointer}

We first study a pointer having one degree of
freedom (it contains only one particle moving in 1-dimension space), which allows us to introduce the model and the notation.

\subsection{Wave functions}

\begin{figure}[th]
\par
\begin{center}
\includegraphics[trim = 0mm 0mm 0mm -10mm,clip,width=9cm]{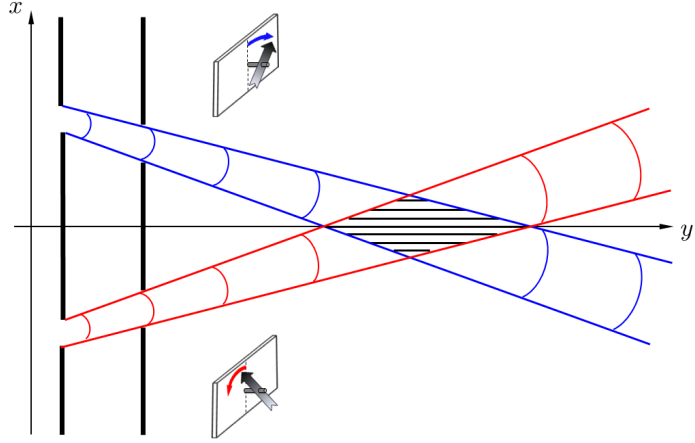}
\end{center}
\caption{Scheme of the experiment discussed in the text. The wave function of
a test particle crosses the two slits of a screen, and then propagates to a region
of space where interferences may take place. The horizontal lines in this
region schematize the positions of the fringes. A pointer particle plays the
role of a Welcher Weg apparatus. When the test particle crosses the upper
slit, the pointer starts moving in one direction, but in the opposite direction when the test
particle crosses the lower slit. The pointer may be either fast (it indicates
a result before the test particle reaches the interference region), slow (it
indicates a result only after the test particle has crossed this region), or intermediate.}%
\label{Fig-1}%
\end{figure}

The test particle is assumed to move in a 2-dimension space, with a wave
function depending on coordinates $x$ and $y$, as shown schematically in
Figure \ref{Fig-1}. The coordinate variable of the 1-dimension pointer is $z$. Just
after the test particle has crossed the screen pierced with two slits, it is
entangled with the variable of the pointer; the total wave function is the sum
of two components:
\begin{equation}
\Psi(x,y,z;t)=\Phi_{+}(x,y,z;t)+\Phi_{-}(x,y,z;t) \label{art-1}%
\end{equation}
where $t$ is the time. In (\ref{art-1}), each component is the product of a
wave function for the test particle (which is itself a product of functions
$\varphi_{\pm}^{x}$ of $x$ and $\varphi^{y}$ of $y$), and of a wave function
for the pointer $\chi_{\pm}$ containing the $z$ dependence:%
\begin{equation}
\Phi_{\pm}(x,y,z;t)=\varphi_{\pm}^{x}(x;t)~\varphi^{y}(y;t)~\chi_{\pm}(z;t)
\label{art-1-1}%
\end{equation}
Note that both components of (\ref{art-1}) contain the same function
$\varphi^{y}$: for the sake of simplicity, we assume that the motion of the
test particle in direction $Oy$ (perpendicularly to the screen) is independent
of the slit crossed by the particle.

Functions $\varphi$ and $\chi$\ are assumed to be simple Gaussian wave packets
(Gaussian slits):%
\begin{equation}
\varphi_{\pm}^{x}(x;t)\sim\left[  a^{4}+\frac{4\hslash^{2}t^{2}}{m^{2}%
}\right]  ^{-1/4}\exp\left\{  \mp i\frac{mv_{x}x}{\hslash}-\frac{\left[  x\mp
d\pm v_{x}t\right]  ^{2}}{a^{2}+\frac{2i\hslash t}{m}}\right\}
\label{art-1-2}%
\end{equation}
which at time $t=0$ is centered at a position $x=\pm d$, and has initial
velocity $\mp v_{x}$ (the signs are chosen so that the wave packets of the
test particle cross in the interference region if $v_{x}>0$); $m$ is the mass of the test particle.
For the second coordinate $y$ of the test particle, we set:%
\begin{equation}
\varphi^{y}(y;t)\sim\left[  b^{4}+\frac{4\hslash^{2}t^{2}}{m^{2}}\right]
^{-1/4}\exp\left\{  i\frac{mv_{y}y}{\hslash}-\frac{\left[  y-v_{y}t\right]
^{2}}{b^{2}+\frac{2i\hslash t}{m}}\right\}  \label{art-1-3}%
\end{equation}
which has an initial width $b$, is centered at $y=0$, and has an initial
velocity $v_{y}$. Finally, for the wave function of the pointer, we choose:%
\begin{equation}
\chi_{\pm}(z;t)\sim\left[  c^{4}+\frac{4\hslash^{2}t^{2}}{M^{2}}\right]
^{-1/4}\exp\left\{  \pm i\frac{MVz}{\hbar}-\frac{\left[  z\mp Vt\right]  ^{2}%
}{c^{2}+\frac{2i\hslash t}{M}}\right\}  \label{art-1-4}%
\end{equation}
where $M$ is the mass of the pointer particle. This wave packet has an initial width $c$ and an initial velocity $+V$ if the test particle crosses the upper slit, $-V$ if it crosses the lower slit (with this convention, the signs of the velocities of the wave packets associated with the test and pointer particles are opposite if $v_x$ and $V$ are both positive). This corresponds to what Ref. \cite{Vaidman-2012}  calls a \textquotedblleft Bohmian velocity detector\textquotedblright.

\subsection{Motion of the Bohmian positions}

We now introduce Bohmian positions $X$ and $Y$ for the test particle, and $Z$ for
the pointer particle. The guiding equation of the Bohmian position of the test
particle reads:
\begin{align}
\frac{\text{d}X}{\text{d}t}  &  =\frac{\hslash}{2im\left\vert \Psi\left(
X,Y,Z\right)  \right\vert ^{2}}\left[  \Psi^{\ast}\left(  X,Y,Z\right)
\frac{\partial \Psi}{\partial x}\left(  X,Y,Z\right)  -\text{c.c.}\right]
\nonumber\\
\frac{\text{d}Y}{\text{d}t}  &  =\frac{\hslash}{2im\left\vert \Psi\left(
X,Y,Z\right)  \right\vert ^{2}}\left[  \Psi^{\ast}\left(  X,Y,Z\right)
\frac{\partial \Psi}{\partial y}\left(  X,Y,Z\right)  -\text{c.c.}\right]
\label{art-4}%
\end{align}
(c.c. means \textquotedblleft complex conjugate\textquotedblright; the time
dependence is not explicitly written for simplicity).\ Equivalently, this Bohmian velocity can also be written in terms of the gradient of the phase
of the wave function. As for the motion of
the pointer Bohmian position, it is given by:%
\begin{equation}
\frac{\text{d}Z}{\text{d}t}=\frac{\hslash}{2iM\left\vert \Psi\left(
X,Y,Z\right)  \right\vert ^{2}}\left[  \Psi^{\ast}\left(  X,Y,Z\right)
\frac{\partial \Psi}{\partial z}\left(  X,Y,Z\right)  -\text{c.c.}\right]
\label{art-5}%
\end{equation}

Since the wave function is factorized with respect to the $y$ variable, the
Bohmian motion along axis $Oy$ is independent of the other variables: there is
actually no $X$ or $Z$ dependence in the second equation (\ref{art-4}). The
motion of $Y(t)$ is therefore given by:%
\begin{equation}
\frac{\text{d}Y}{\text{d}t}=v_{y}+\frac{4\hbar^{2}t}{m}\frac{Y-v_{y}t}%
{b^{4}+\frac{4\hbar^{2}t^{2}}{m^{2}}} \label{art-11}%
\end{equation}
The solution of this equation is:%
\begin{equation}
Y(t)=v_{y}t+Y_{0}\sqrt{1+\frac{4\hbar^{2}t^{2}}{m^{2}b^{4}}} \label{art-12}%
\end{equation}
where $Y_{0}$ is the initial value of $Y$ at time zero. The motion of $Y(t)$
is therefore relatively simple: a uniform motion along $Oy$ plus a correction
introduced by the diffraction of the wave packet.

\subsection{Dimensionless variables}

We now introduce dimensionless variables by setting:%
\begin{align}
x^{\prime}  &  =\frac{x}{a}~~~~~~~~~y^{\prime}=\frac{y}{b}~~~~~~~~~~~z^{\prime
}=\frac{z}{c}\nonumber\\
X^{\prime}  &  =\frac{X}{a}~~~~~~~~Y^{\prime}=\frac{Y}{b}~~~~~~~~~Z^{\prime
}=\frac{Z}{c} \label{art-6}%
\end{align}
and:%
\begin{equation}
d^{\prime}=\frac{d}{a} \label{art-6-2}%
\end{equation}
All positions are then expressed in terms of the initial width of the
corresponding wave packet.\ We also introduce a dimensionless time $t^{\prime
}$ by:%
\begin{equation}
t^{\prime}=\frac{v_{y}t}{b} \label{art-7}%
\end{equation}
The velocities are defined in terms of the variables:%
\begin{equation}
\xi_{x}=\frac{mv_{x}a}{\hslash}~~~~~~~~~~~~~~~~\xi_{y}=\frac{mv_{y}b}{\hslash
}~~~~~~~~~~~~~~~~\Xi=\frac{MVc}{\hbar} \label{art-8}%
\end{equation}
Equation (\ref{art-12}) then simplifies into:%
\begin{equation}
Y^{\prime}(t^{\prime})=t^{\prime}+Y_{0}^{\prime}\sqrt{1+\frac{4(t^{\prime
})^{2}}{\xi_{y}^{2}}} \label{art-12-bis}%
\end{equation}

Finally, it is also convenient to introduce the dimensionless parameters:%
\begin{equation}
r=\frac{a}{b}~~~~~~~~~~~~~~~~R=\frac{a}{c}~~~~~~~~~~~~~~~~\mu=\frac{m}{M}
\label{art-13}%
\end{equation}
The equations of motion of the Bohmian positions of the test and pointer
particles then become:%
\begin{align}
\frac{\text{d}X^{\prime}}{\text{d}t^{\prime}}  &  =\frac{1}{2ir^{2}\xi
_{y}\left\vert \Psi\right\vert ^{2}}\left[  \Psi^{\ast}\frac{\partial
}{\partial x^{\prime}}\Psi-\Psi\frac{\partial}{\partial x^{\prime}}\Psi^{\ast
}\right] \nonumber\\
\frac{\text{d}Z^{\prime}}{\text{d}t^{\prime}}  &  =\frac{\mu R^{2}}{2ir^{2}%
\xi_{y}\left\vert \Psi\right\vert ^{2}}\left[  \Psi^{\ast}\frac{\partial
}{\partial z^{\prime}}\Psi-\Psi\frac{\partial}{\partial z^{\prime}}\Psi^{\ast
}\right]  \label{art-14}%
\end{align}

The initial values of the Bohmian positions \ are denoted as $X_{0}$, $Y_{0}$
and $Z_{0}$ respectively (with added primes for dimensionless variables).\ For a given realization of the experiment, these
variables are random, with a distribution given by the squared modulus of the
corresponding wave function.\ For instance, $X_{0}$ can fall at any place
inside one of the two Gaussian slits; for all figures of this article, we show the
trajectories corresponding to $9$ equidistant values inside each slit.\ Since
$Y$ does not play an important role in the problem (it just increases smoothly
in time), we set $Y_{0}=0$, so that $Y$ (or $Y^{\prime}$) provides a rough measure of the
time.\ Finally, to study the impact of the values of $Z_{0}$ on the
trajectories, we make a random choice of $Z_{0}$ inside its Gaussian distribution.

The time at which the wave packets of the test particle cross in the interference region is
$t_{\text{cross}}\simeq d/v_{x}$, and the time at which the wave packets of
the pointer become well separated is $t_{P}\simeq c/V$. The pointer thus gives a clear
indication before the test particle reaches the interference region if:%
\begin{equation}
\frac{t_{\text{cross}}}{t_{P}}\simeq\frac{Vd}{v_{x}c}>1 \label{art-14-2}%
\end{equation}
In terms of the dimensionless variables introduced above, this condition
becomes:%
\begin{equation}
E=\frac{Vd}{v_{x}c}=\frac{\Xi}{\xi_{x}}R^{2}d^{\prime}\mu>1 \label{art-14-3}%
\end{equation}
For short, we will call (\ref{art-14-3}) the \textquotedblleft fast pointer
condition\textquotedblright.


Bohmian trajectories may be obtained by straightforward time integration of the appropriate set of coupled differential equations: Eq.~(\ref{art-14}) for a single-particle pointer, a set of $1+N$ similar equations for the Bohmian positions associated to the test particle,  $X'$,  and to the multi-particle pointer(s), $Z'_{1}$ to $Z'_{N}$, see Appendix. With conventional numerical tools or dedicated software, this is manageable on a personal computer for reasonably small values of $N$ (and, possibly, any type of wave packets).

For our Gaussian two-slit model, we have used Mathematica \cite{Mathematica} for computation and plotting of all trajectories. Solving this type of ordinary differential equations with Mathematica's generic tool and default options works well up to a few tens of pointer particles and time integration is fast (a fraction of a second up to  $N=5$, typically). For larger $N$ values, we resorted to the (automatically) suggested optional method of simplification suited for the problem, which is more robust but slower. The computing time increases exponentially with the number of pointer particles, ranging from 5~sec. for $N=10$ to 3.5~hours for $N=200$ on a computer equipped with a 3.2~GHz Intel Core i3-550 CPU, for example, or 1.5~h for $N=200$ on a 4.0~GHz Intel Core  i7-6700K CPU.

One way to substantially reduce the computational load is to use the explicit analytical expressions of the Bohmian velocities, which can be derived for Gaussian wave packets (see Appendix, Section 2). Then a roughly 10-fold shorter time is needed for computation of all trajectories.

Finally and more simply, as shown in Section 3 of the Appendix, the particle and pointer trajectories may actually be obtained from the time integration of only 2 differential equations. These two equations rule the evolution of $X^{\prime}$ and $\hat{\Sigma}$, the Bohmian positions of the test particle and of a single \textquotedblleft new\textquotedblright\ particle associated to the pointer, respectively. The trajectories of all pointer particles can be subsequently derived analytically, from the evolution of a new particle that has an effective velocity parameter $\hat{\Xi} = \Xi\sqrt{N}$ when the pointer includes $N$ particles with (dimensionless) velocity $\Xi$, and therefore moves $\sqrt N$ times faster.

\section{Single particle pointer}

\label{fast-and-slow-pointers}

When the Welcher Weg pointer contains a single particle, the system under
study is made of two entangled particles, exactly as in EPR/Bell non-locality
experiments. Various situations can be considered, depending on the nature of
the entanglement between the test particle and the pointer. For the sake of
simplicity, we first assume that they are not coupled, a case in which no
Welcher Weg information whatsoever is provided by the pointer; it
has no effect on the trajectory of the test particle.\ Next we consider the
opposite situation, where the pointer is strongly coupled to the test particle and and provides a Welcher Weg information before the test particle reaches the interference region (fast pointer); the trajectories are
then completely different and look classical.\ We also study what happens if the pointer is slow
and provides the information only after the test particle has crossed the
interference region, and finally consider intermediate situations.

\subsection{Immobile pointer providing no information}

\begin{figure}[tbh]
\begin{center}
\includegraphics[width=7cm]{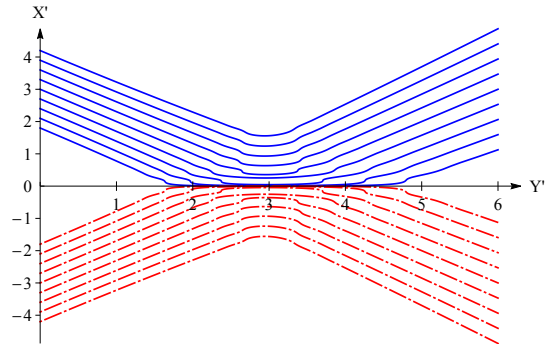}
\end{center}
\caption{(color on line) Trajectories obtained when the pointer particle is not coupled to the
test particle ($\Xi=V=0$). The two particles are not entangled and their positions move independently.\ The figure shows
the trajectories of the test particle originating from $9$ possible initial
values $X_{0}$ inside each of the slits.\ The trajectories crossing the upper slit are drawn with full lines (blue color on line), those crossing
the lower slit in dashed lines (red color on line). All trajectories of the test particle
\textquotedblleft bounce\textquotedblright\ on the symmetry plane of the
experiment (no-crossing rule).}%
\label{Fig-2}%
\end{figure}

If we set $\Xi=E=0$, the wave function (\ref{art-1}) factorizes into
a component for the test particle and another for the pointer, so that the motions
of their Bohmian positions become independent: the motion of
the test particle occurs as if the apparatus did not exist.\ Figure
\ref{Fig-2} shows the trajectories obtained in this case. They exhibit the
so-called \textquotedblleft no-crossing rule\textquotedblright: trajectories
cannot cross the symmetry plane of the experiment, and they \textquotedblleft
bounce\textquotedblright\ on this plane in the interference region. This is a simple
consequence of the fact that, in the symmetry plane, the probability current
is contained inside the plane.\ In dBB quantum mechanics, it is well-known
that interference effects can curve trajectories in free space, so that an observation of the trajectory of the particle after the interference
region should not be extrapolated as a straight line backwards in time.

\subsection{Pointer providing real-time information (fast pointer)}

\begin{figure}[tbh]
\begin{minipage}[c]{0.5\textwidth}
\includegraphics[width=6cm]{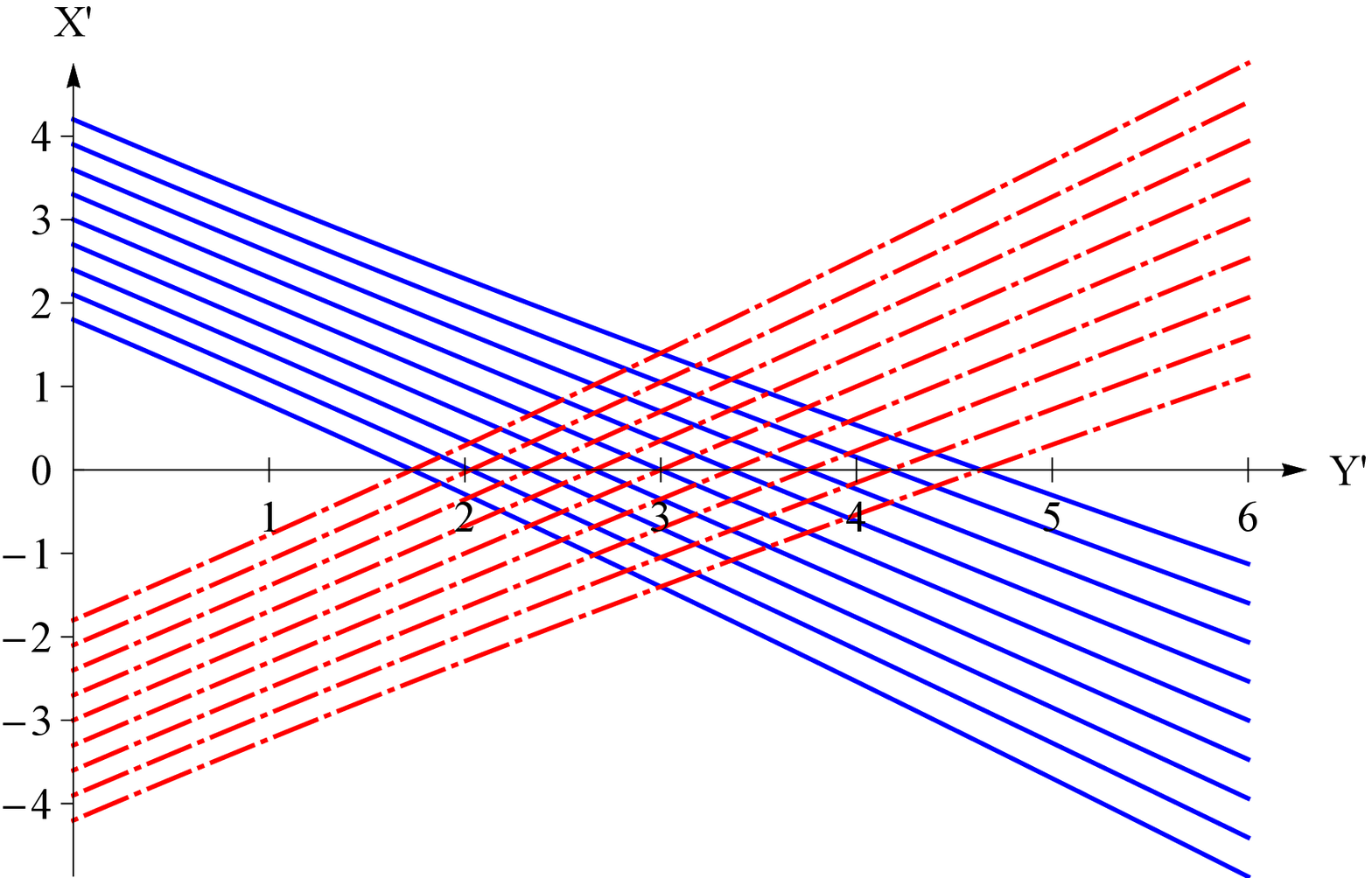}
\end{minipage}\hfill\begin{minipage}[c]{0.5\textwidth}
\includegraphics[width=6cm]{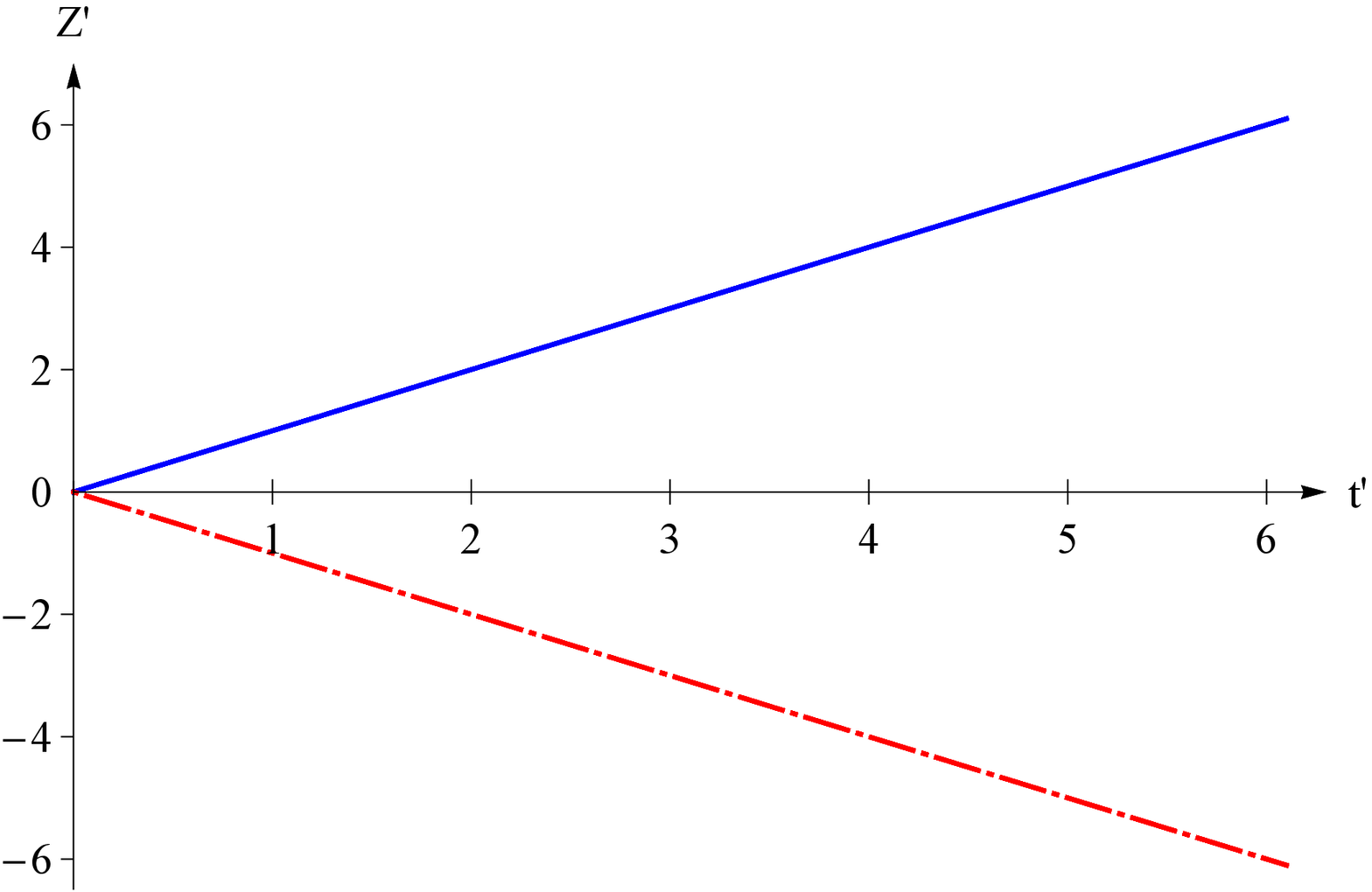}
\end{minipage}
\caption{(color on line) Trajectories obtained when the pointer particle is coupled to the
test particle and plays the role of a \textquotedblleft fast pointer\textquotedblright (i.e. a pointer providing a Welcher Weg information before the test particle
reaches the interference region). The left part of the figure
shows the trajectories of the test particle, the right part the trajectories
of the pointer particle. The values of the input parameters used for the calculation
are $\Xi=\xi_{x}=10$, $r=R=1$, $d^{\prime}=3$, $\mu=1$, corresponding to $E=3$. The position of the pointer particle is assumed to be initially centered ($Z_0^{\prime} = 0$). It subsequently moves upwards if the test particle crosses the upper slit, downwards if it crosses the lower slit. Because this information may be obtained before the test particle reaches the interference region, no interference effect takes place, so that the trajectories of the test particle
are (almost) straight lines. Similarly, the trajectories of the pointer particle also remain straight, and depend only on the slit crossed by the test particle, independently on its position inside the slit. No trajectory
can then be interpreted as \textquotedblleft surrealistic\textquotedblright .}%
\label{Fig-3}%
\end{figure}

We now assume that the pointer is sufficiently coupled to the test particle to
provide a real-time information: it indicates the slit crossed by the particle
before the interference region is reached. This situation is obtained when
condition (\ref{art-14-3}) for $E$ is met.\ As shown in Figure \ref{Fig-3},
the trajectories then cross the interference region as almost straight
lines.\ This was expected since, in quantum mechanics, any Welcher Weg
information stored in the pointer necessarily destroys the interference
effects.\ One can then infer which slit was crossed by the test particle by
a simple backwards extrapolation of its trajectory along a straight line; no surrealistic trajectory appears.

\subsection{Pointer providing only delayed information (slow pointer)}

\begin{figure}[tbh]
\begin{minipage}[c]{0.5\textwidth}
\includegraphics[width=6cm]{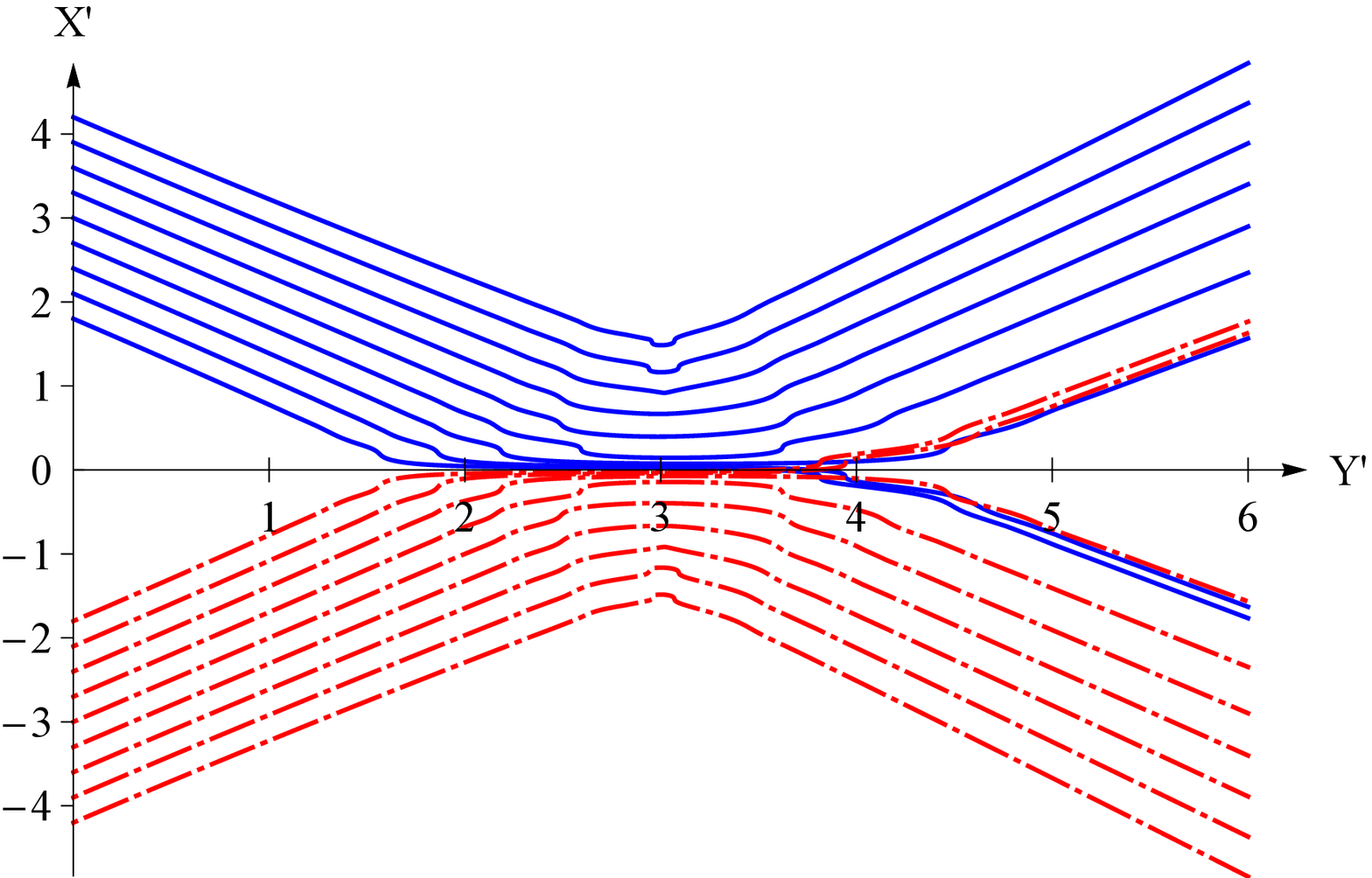}
\end{minipage}\hfill\begin{minipage}[c]{0.5\textwidth}
\includegraphics[width=6cm]{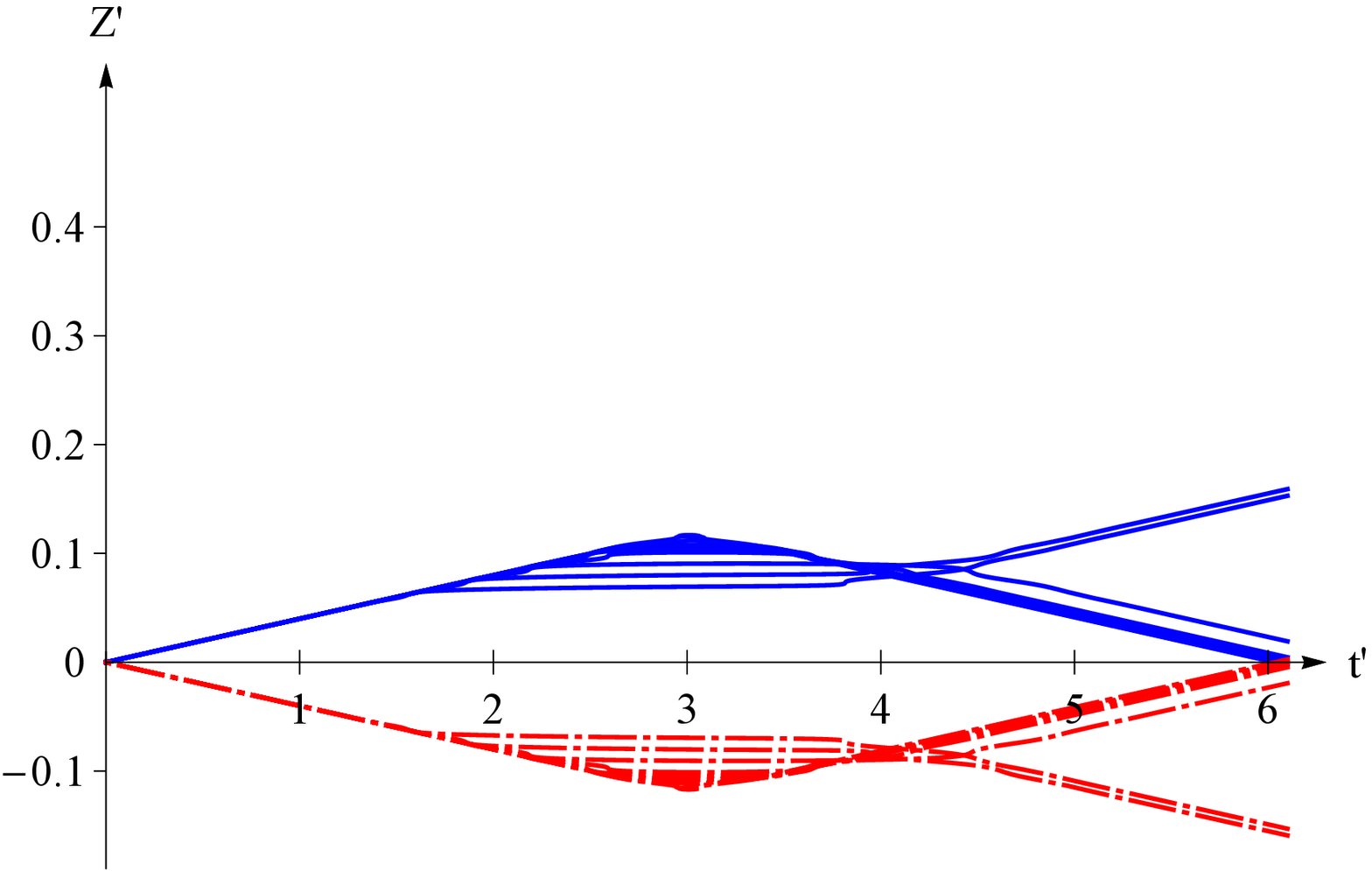}
\end{minipage}
\caption{(color on line) Trajectories obtained when the pointer particle is slow: it provides
only a delayed Welcher Weg information after the test particle has crossed the
interference region. Interference effects can then occur for
the test particle. The only input parameter that has been changed with respect to
Figure \ref{Fig-3} is $R=0.2$ (this corresponds to $E=0.12$). Most trajectories now change direction during the
crossing of the interference region by the test particle. These curved trajectories  are sometimes called
\textquotedblleft surrealistic trajectories\textquotedblright\ because the position of the pointer, long after the crossing, has a behavior opposite of that of a fast pointer: it moves downwards if the trajectory of the test particle crosses the upper slit, upwards if it crossed the lower slit. A few trajectories remain \textquotedblleft normal\textquotedblright\
and do not exhibit this apparent contradiction. In this figure, we have assumed that the
initial Bohmian position of the pointer particle vanishes ($Z_{0}^{\prime} =
0$); the trajectories, thus, form a perfectly symmetric
pattern.}%
\label{Fig-4}%
\end{figure}

We now assume that the pointer is slow: its wave packets separate
significantly only after the test particle has crossed the interference
region. This situation corresponds to what Ref. \cite{Dewdney-et-al-1993}
calls \textquotedblleft late measurements\textquotedblright.
Figure~\ref{Fig-4} is obtained by changing parameter $R$
from $R=1$ (Figure~\ref{Fig-3}) to $R=0.2$. At constant $\Xi$, this reduces the velocity of the pointer
by a factor $5$; moreover, at constant $a$ (no change of the wave packet of the test particle), this multiplies the width of the wave packet of the pointer
particle by a factor $5$; altogether, parameter $E$ is divided by $25$ and becomes equal to $E=0.12$.

We then obtain a completely different situation. Initially,
the pointer starts to move in a direction that indicates the slit crossed
by the Bohmian position of the test particle, as expected (and as in Figure \ref{Fig-3}). But, when
the test particle reaches the interference region, non-local interference effects take
place: the Bohmian positions can jump\footnote{We use the word \textquotedblleft jump\textquotedblright\ as several other authors have done, because it is convenient. This does not mean that there is a real jump is space, since the Bohmian trajectory remains perfectly continuous; one could also say that the Bohmian positions change the wave on which they surf.} from one wave packet to the other, so that
both trajectories change directions. They remain consistent with each
other, inasmuch as the motion of the pointer particle (its velocity) constantly indicates the beam in which
the test particle propagates.

Nevertheless, if one extrapolates the behavior of a fast pointer to this case, one reaches a contradiction. If for instance the test particle crossed the upper slit,
at long times the pointer is more likely to move downwards (as a fast pointer would do if the test particle had crossed the lower slit, see Fig. \ref{Fig-3}). Ref.~\cite{Englert-1992} considers that this motion of the pointer provides a correct information on which slit was really crossed by the test particle; because the Bohmian trajectory crossed the other slit, it becomes contradictory with (this interpretation of) the measurement, therefore physically meaningless and  \textquotedblleft surrealistic\textquotedblright .

\begin{figure}[!tb]
\begin{minipage}[c]{0.5\textwidth}
\includegraphics[width=6cm]{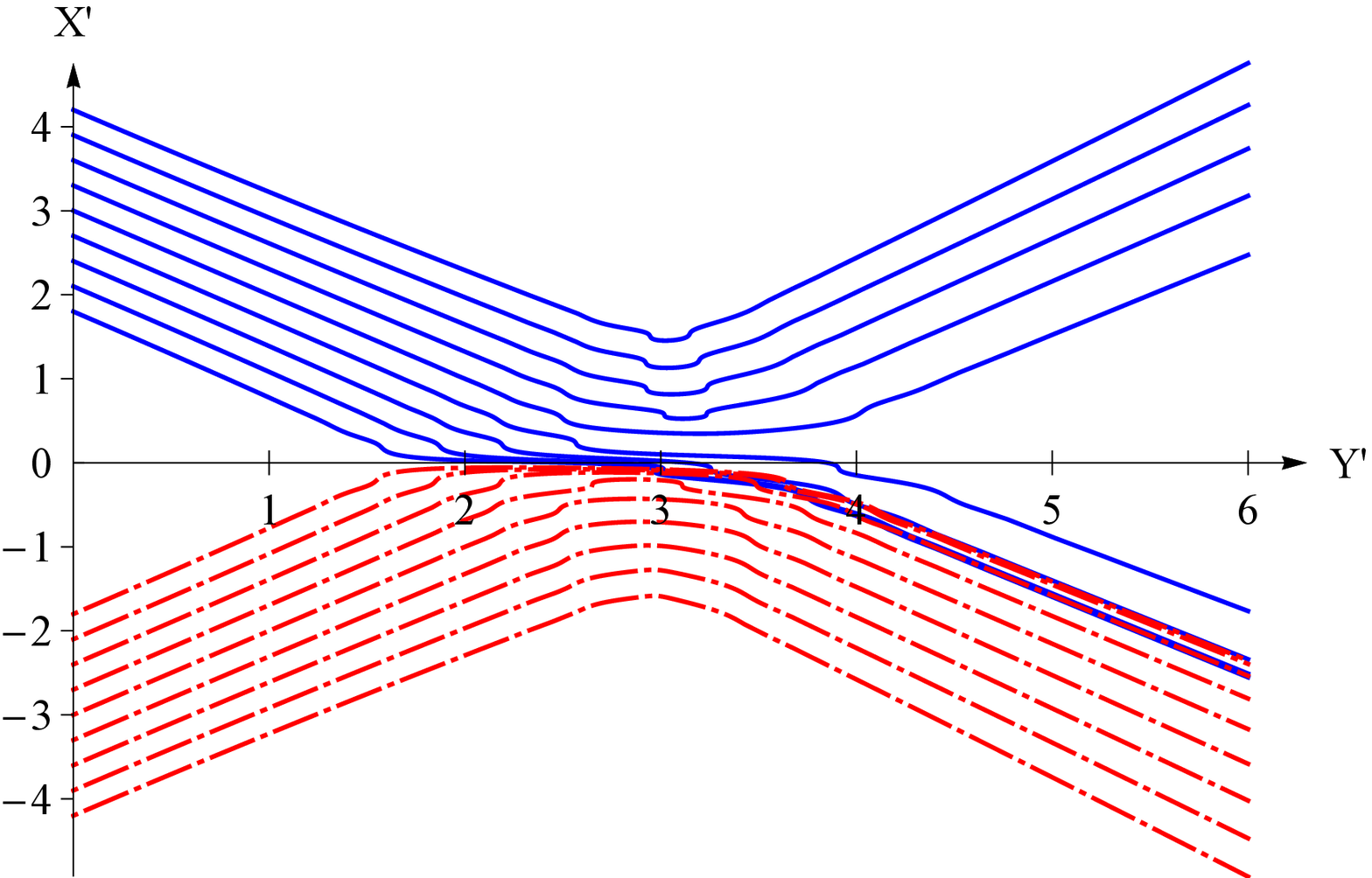}
\end{minipage}\hfill\begin{minipage}[c]{0.5\textwidth}
\includegraphics[width=6cm]{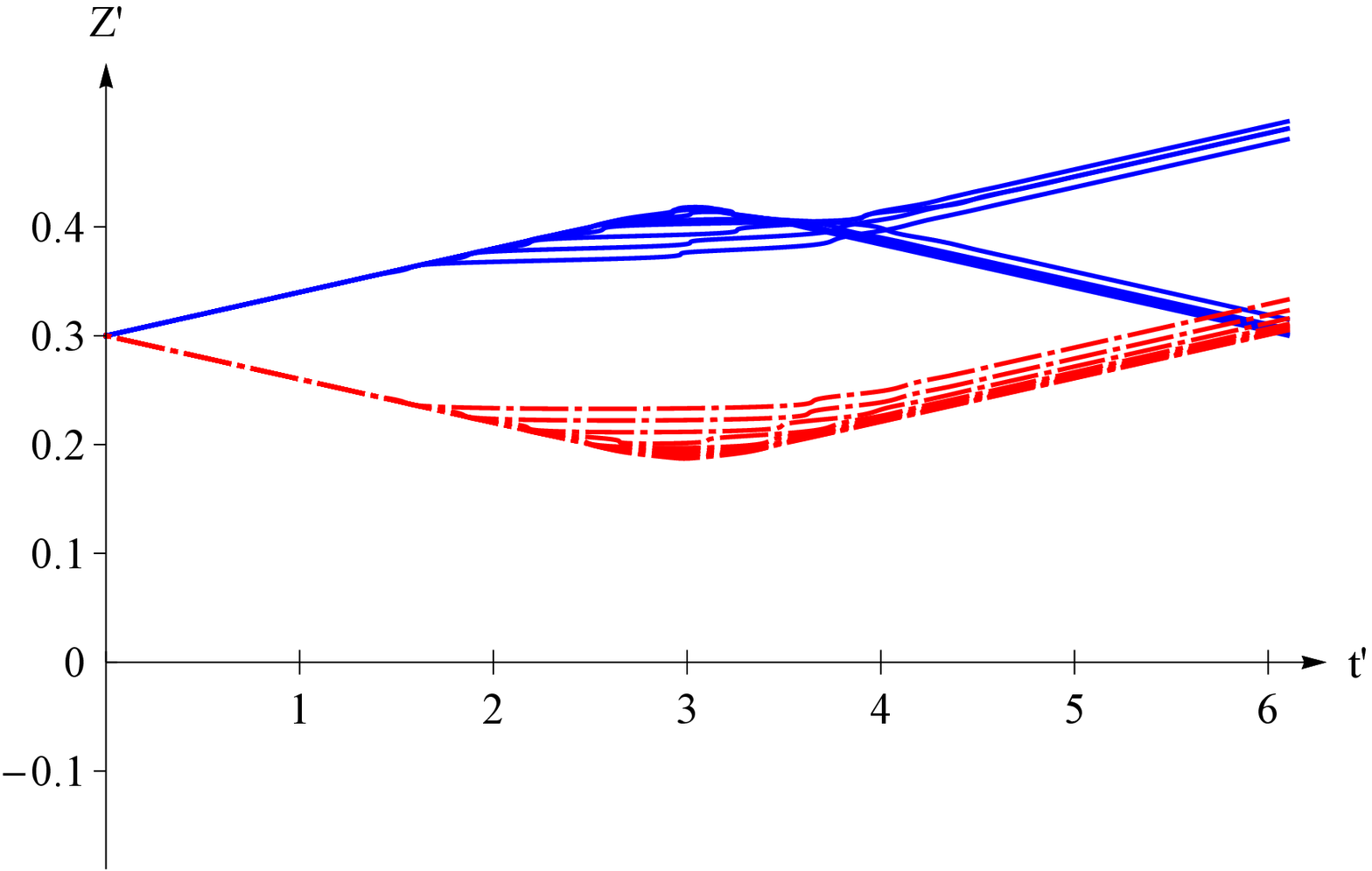}
\end{minipage}
\caption{(color on line) Same figure as Fig.~\ref{Fig-4}, but $Z_{0}^{\prime}$ has been changed to
$Z_{0}^{\prime}=0.3$, which introduces an asymmetry. The proportion of situations where
the pointer position ends up moving upwards (and the test particle downwards) has increased. This illustrates how the initial value of an \textquotedblleft additional variable\textquotedblright\ of the measurement apparatus can
influence the future trajectory of the measured system, due to a quantum non-local
effect. After the interference region, the trajectory of the test particle takes more often a
direction that seems to originate from the slit corresponding to the initial
indication of the pointer (\textquotedblleft predestination
effect\textquotedblright).}%
\label{Fig-5}%
\end{figure}

Another interesting feature of this situation is that the initial value $Z_{0}$ of the Bohmian position of the pointer
plays an important role. It is a random variable with a probability
distribution that is determined by the squared modulus of the initial wave
function $\chi_{\pm}(z;t=0)$. Figure~\ref{Fig-4} corresponds to the case where
$Z_{0}=0$, when the initial position of the pointer is perfectly centered, so
that no slit is favored. In Figure~\ref{Fig-5}, the dimensionless value of $Z_{0}$ has been changed to
$Z_{0}^{\prime}=0.5$; we see that the trajectories are significantly different. We
actually observe an interesting non-local \textquotedblleft
predestination\textquotedblright\ effect: because the initial
position of the pointer particle seems to already point to one of the slits, in the interference region
the trajectory of the test particle is influenced by this indication. We have computed series of trajectories for numerous random initial values of the position of the pointer particle. Systematically, the most frequently selected direction seems to originate from the slit selected by the pointer from the beginning; the
measured system \textquotedblleft follows\textquotedblright\ the initial indications of
the measurement apparatus, so to say.

\begin{figure}[!tb]
\begin{center}
\includegraphics[width=8cm]{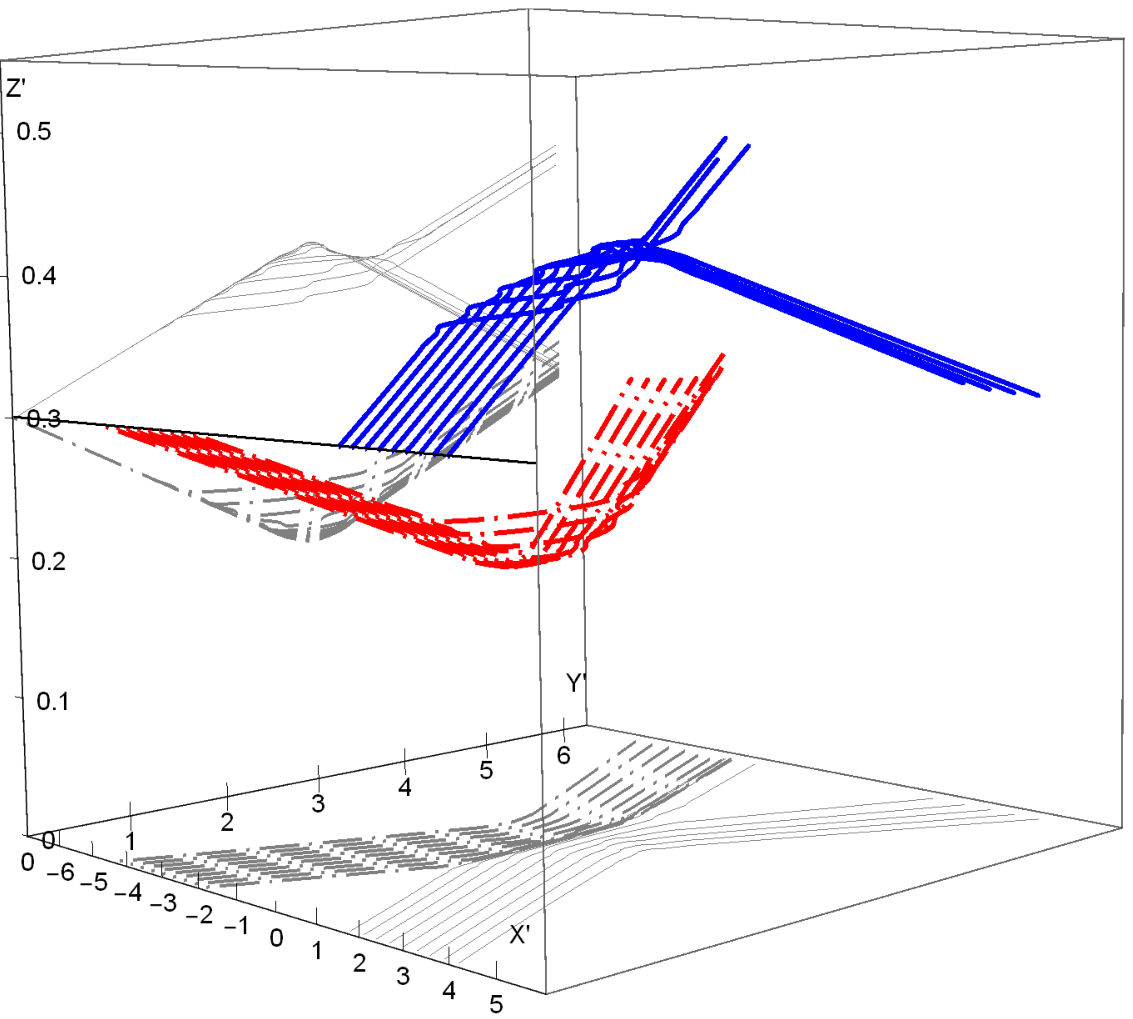}
\end{center}
\caption{(color on line) Same figure as Fig.~\ref{Fig-5}, but in a three dimension representation showing the simultaneous motion of the test and pointer particle. This illustrates how the trajectories avoid each other in the third dimension. 2D projections, showing the trajectories of the test particle and of the pointer, are also plotted in thin lines and gray color.}%
\label{Fig-6}%
\end{figure}

It is known that Bohmian trajectories can never cross each other. The apparent crossings of Figure \ref{Fig-5} are due to the fact that this figure is a projection over a 2D plane of a trajectory that actually takes place in a 3D space, the third dimension being the position of the pointer particle. Figure \ref{Fig-6} illustrates how the trajectories avoid crossing in the third dimension.

This brief survey of the behavior of a microscopic pointer shows that the random initial
position of the pointer plays an important role, and may even influence the
 trajectory of the test particle.\ The possible
\textquotedblleft surrealistic\textquotedblright\ character of the trajectory
is interpretation dependent; surrealism appears only if one interprets the
indications of the pointer as providing information about the past positions
of the test particle.\ Nevertheless, at any time the motion of the pointer
particle gives an information that is consistent with the present motion of
the test particle: just after it crossed the slit, the position of the pointer
moves in the corresponding direction; if the test particle later jumps from
one wave packet to the other, the pointer also reverses its motion, providing
an indication that remains consistent with the present trajectory of the test
particle.\ Nothing seems surrealistic if the non-local dynamics of the measurement process
is duly taken into account.

\section{One or two pointers containing several particles}
\label{pointer-several-particles}

\subsection{Description of the model}

We now generalize (\ref{art-1-1}) by assuming that each of the two components
of the wave function contains the product of $N$ individual wave functions
associated with particles contained in one or several pointers.\ With the
same dimensionless variables as defined above, we write:%
\begin{align}
\Phi_{\pm}(x,y,z_{1},z_{2,..,}z_{N};t)  &  \propto\nonumber\\
&  ~~\exp\left\{  -\frac{\left[  x^{\prime}\mp d^{\prime}\pm\frac{\xi_{x}%
}{r^{2}\xi_{y}}t^{\prime}\right]  ^{2}}{1+\frac{4t^{\prime2}}{r^{4}\xi_{y}%
^{2}}}-\frac{\left[  y^{\prime}-t^{\prime}\right]  ^{2}}{1+\frac{4t^{\prime2}%
}{\xi_{y}^{2}}}-\sum_{n=1}^{N}\frac{\left[  z_{n}^{\prime}  - \frac{\mu\Xi_n^{\pm} \,
R^{2}}{r^{2}\xi_{y}}t^{\prime}\right]  ^{2}}{1+4\frac{\mu^{2}R^{4}t^{\prime2}%
}{r^{4}\xi_{y}^{2}}}\right\} \nonumber\\
&  \times\exp\left\{  i\left[  \left(  \pm\xi_{x}x^{\prime}+\xi_{y}y^{\prime
} + \sum_{n=1}^{N} \Xi_n^{\pm} \,  z_{n}^{\prime}~\right)  +\frac{2t^{\prime}}{r^{2}\xi_{y}%
}\frac{\left[  x^{\prime}\mp d^{\prime}\pm\frac{\xi_{x}}{r^{2}\xi_{y}%
}t^{\prime}\right]  ^{2}}{1+\frac{4t^{\prime2}}{r^{4}\xi_{y}^{2}}}\right.
\right. \nonumber\\
&  ~~~~~~~~~~~~~~~~~\left.  \left.  +\frac{2t^{\prime}}{\xi_{y}}%
\frac{\left[  y^{\prime}-t^{\prime}\right]  ^{2}}{1+\frac{4t^{\prime2}}%
{\xi_{y}^{2}}}+\sum_{n=1}^{N}\frac{2\mu R^{2}t^{\prime}}{r^{2}\xi_{y}}%
\frac{\left[  z_{n}^{\prime} - \frac{\mu\Xi_n^{\pm} \, R^{2}}{r^{2}\xi_{y}}t^{\prime
}\right]  ^{2}}{1+4\frac{\mu^{2}R^{4}t^{\prime}}{r^{4}\xi_{y}^{2}}}\right]
\right\}  \label{art-100}%
\end{align}
Each $z_{n}$ is the 1D spatial coordinate of one pointer
particle; a different origin may be chosen for each value of $n$, meaning that
the wave packets do not necessarily coincide at time $t=0$. The parameters $\Xi_n^{\pm}$ define the initial
velocities of the wave packets of the pointer particles: if the test particle crosses the upper slit, the (dimensionless) velocity is $\Xi_n^{+}$; if it crosses the lower slit, it is $\Xi_n^{-}$. Of course, if there is only one pointer, all the
pointer particles belong to the same solid object, and we will assume that all these
wave packets move at the same speed; we then simply choose:
\begin{equation}
\Xi_n^{\pm}=\pm \, \Xi
\end{equation}

But we can also assume that two independent  pointers are used to detect the test particle: for instance, one pointer starts moving if this particle crosses the upper slit, but remains still otherwise; the other pointer operates in the same way for the lower slit. The simplest case is obtained if each pointer contains only one particle ($N=2$), and with the following choice of parameters:
\begin{align}
\Xi_1^+ = \Xi \hspace{1cm} & \hspace{1cm} \Xi_1^- = 0  \notag \\ \Xi_2^+=0 \hspace{1cm} & \hspace{1cm} \Xi_2^- = \Xi
\label{article-21}
\end{align}
We now study this latter case.

\subsection{Two independent pointers}
\label{two-pointers}


\begin{figure}[!h]
\begin{center}
\includegraphics[width=7cm]{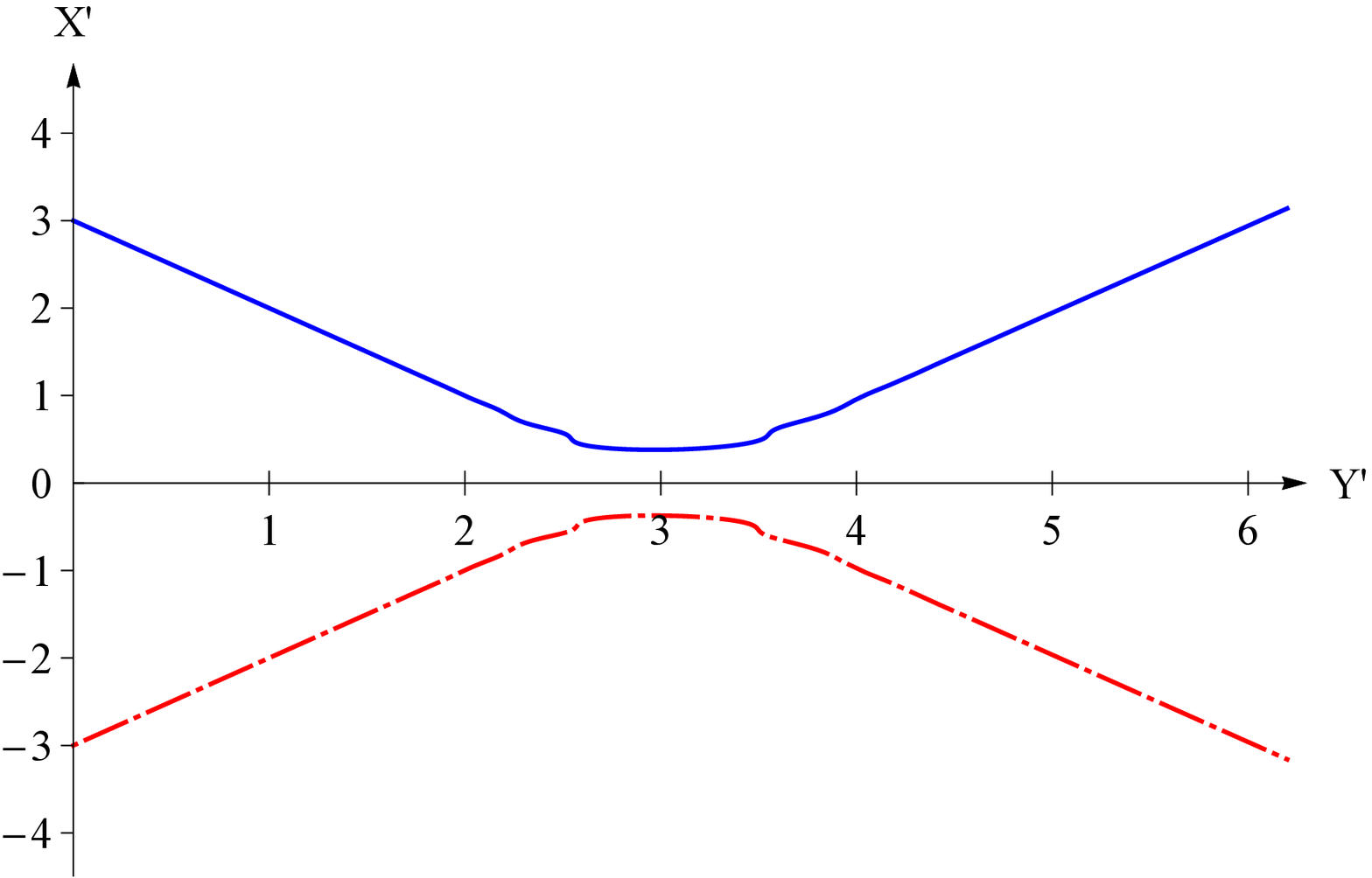}
\end{center}
\begin{minipage}[c]{0.5\textwidth}
\includegraphics[width=6cm]{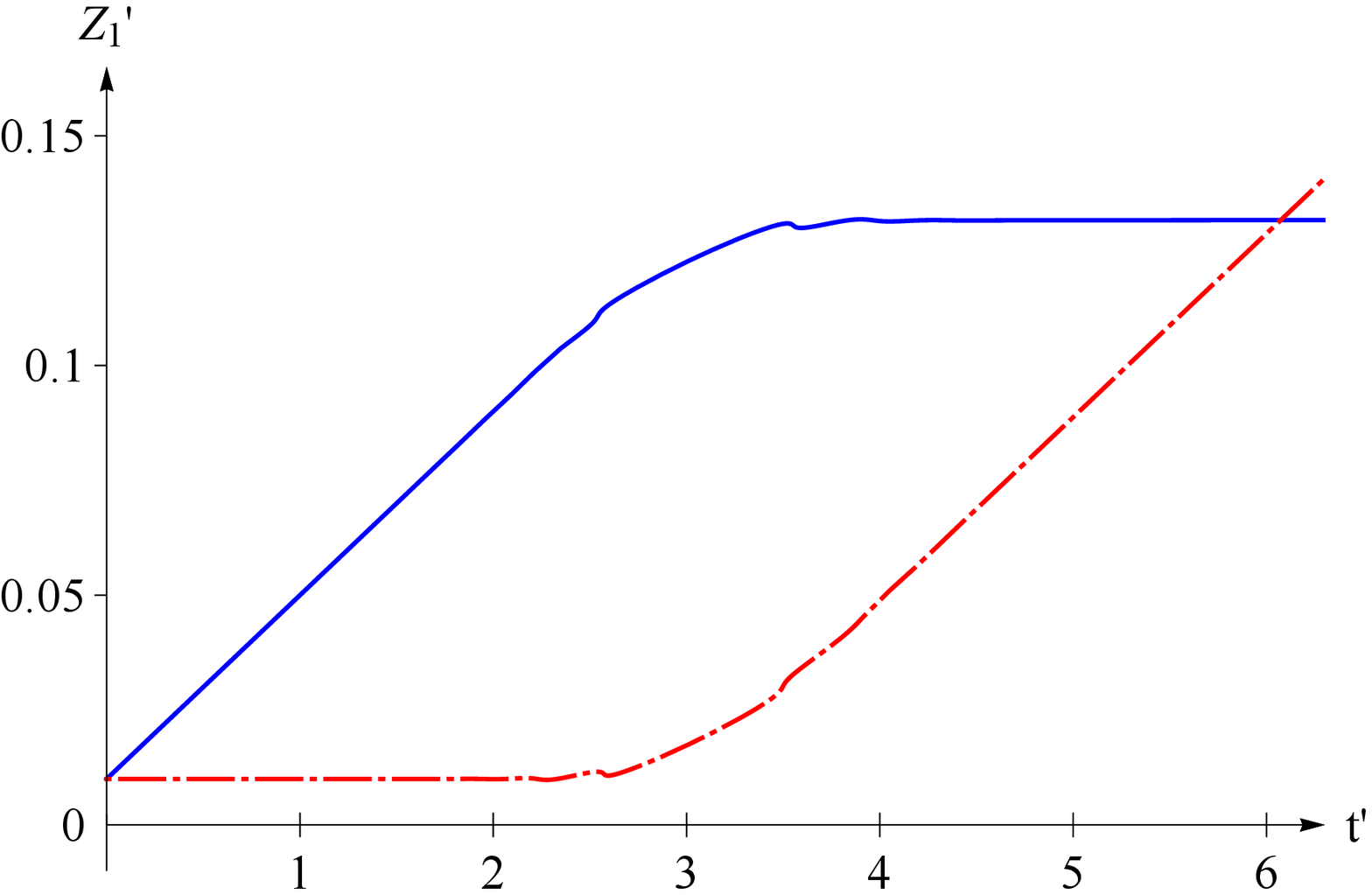}
\end{minipage}\hfill
\begin{minipage}[c]{0.5\textwidth}
\includegraphics[width=6cm]{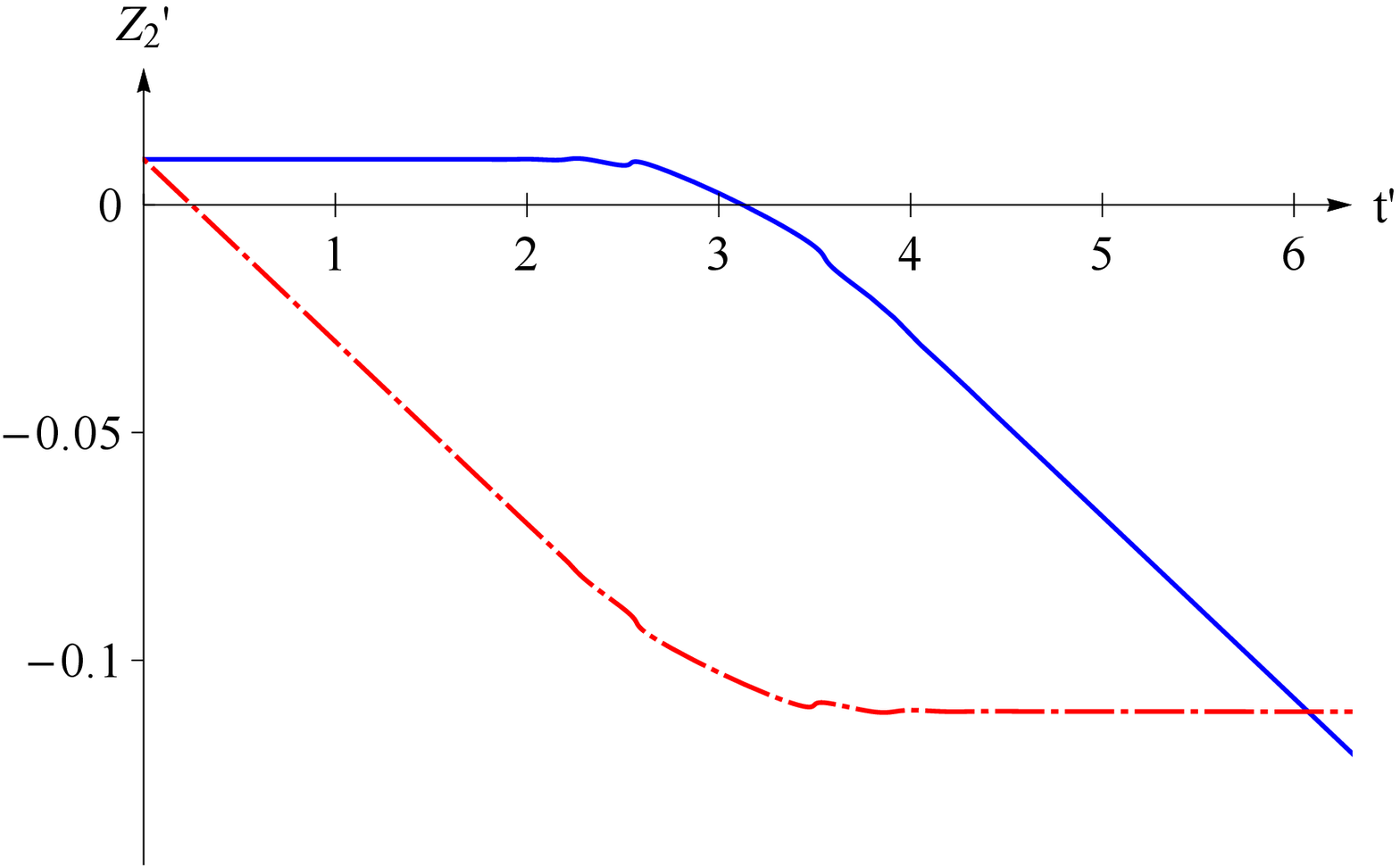}
\end{minipage}
\caption{(color on line) Trajectories obtained with two slow pointers, with the same values of the input
parameters as in Figure \ref{Fig-4}. The initial positions of the pointer particles $Z_{1,0}^{\prime}$ and $Z_{2,0}^{\prime}$ are both equal to $0.01$. The upper part of the figure shows two trajectories of the test particle, one originating from the upper slit (full line, blue on line) and one originating from the lower slit (dashed line, red on line). The trajectories of the two pointers are shown in the lower part  of the figure: on the left the pointer associated with the upper slit, on the right that associated with the lower slit.
\protect{\newline}
Initially, the motions of the positions are exactly those one could naively expect: the position of the pointer located near the slit crossed by the particle moves,  while the other remains still. Later, when the test particle crosses the
interference region, the Bohmian position in the configuration space \textquotedblleft
jump\textquotedblright\ from one wave packet to the other, resulting in the simultaneous
appearance of curved trajectories for each constituent particle. The test particle then changes
direction, the first pointer particle stops, and the second pointer particle
starts to move. This non-local effect, similar to that shown in Figure \ref{Fig-5}, is called \textquotedblleft surrealistic
trajectory\textquotedblright\ in the literature.}%
\label{Fig-7}%
\end{figure}

We consider two microscopic pointers, each made of a single particle with parameters given by (\ref{article-21}).  Figure
\ref{Fig-7} shows two trajectories that are obtained in this case, with the same set of input parameters as for Figures \ref{Fig-4} and \ref{Fig-5}. For clarity, only one trajectory is displayed for each slit, assuming that the test particle crosses it at the center. Initially, when the test particle crosses
the upper slit, the position of the upper pointer moves upwards, while the position of the
lower pointer remains still; this is exactly as expected. Later, when the test
particle reaches the interference region, it can bounce and the velocity can change its
direction; this introduces curvatures in the trajectories of all
three particles. The position of the upper pointer stops, and that of the lower
pointer starts moving up, according to the new
velocity of the test particle --  clearly a quantum non-local effect.
Nevertheless, the successive positions of the pointers give a perfectly correct real-time information on
the actual trajectory of the test particle. No contradiction appears
between the indications of the pointers and the trajectory of the test particle,
provided the motion of the pointers are interpreted as measurements of the
 velocity of the test particle at the same time; this is an interesting illustration of
how a quantum non-local measurement apparatus can operate.

\begin{figure}[t]
\begin{center}
\includegraphics[width=6cm]{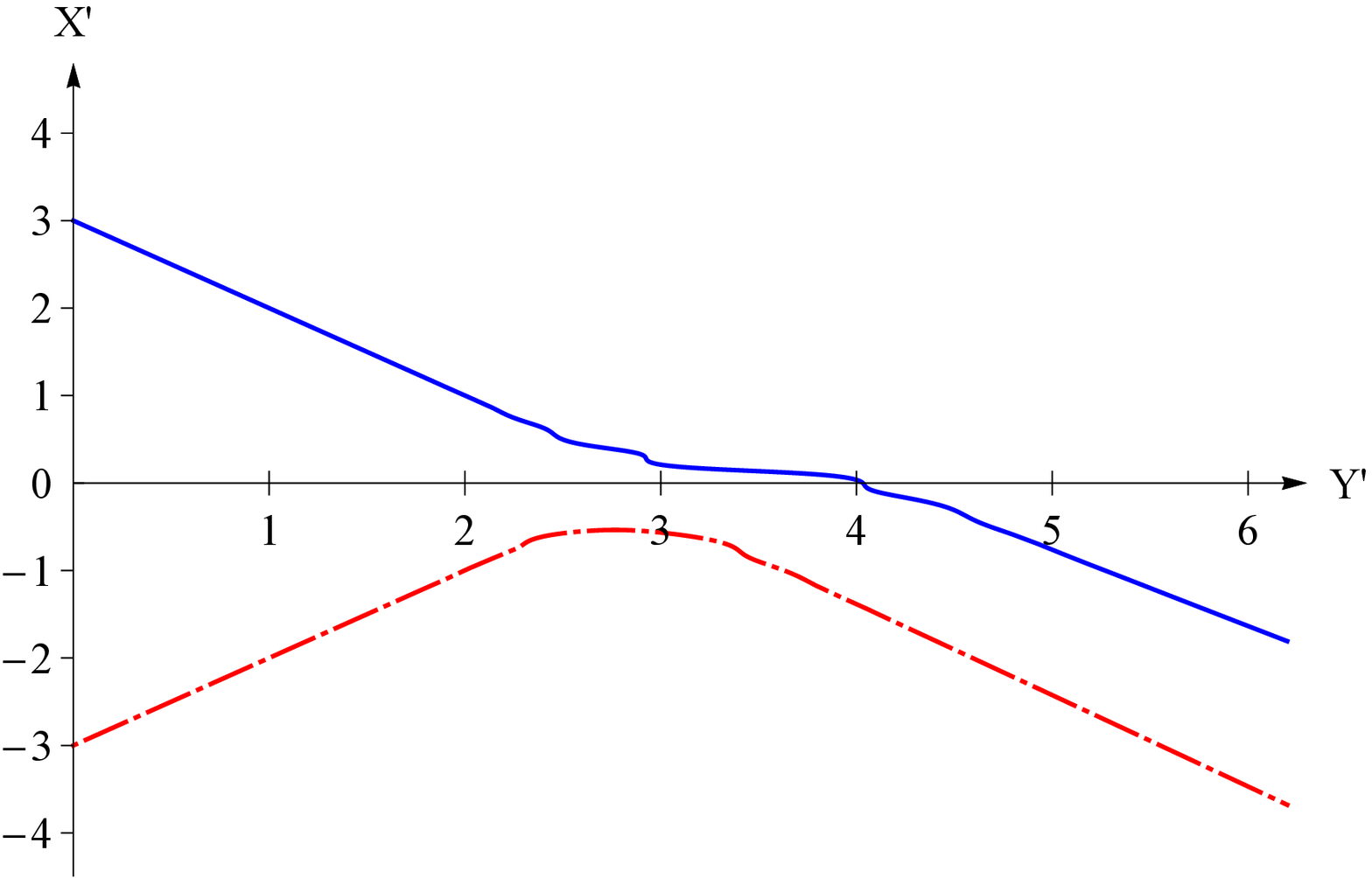}
\end{center}
\begin{minipage}[c]{0.5\textwidth}
\includegraphics[width=5cm]{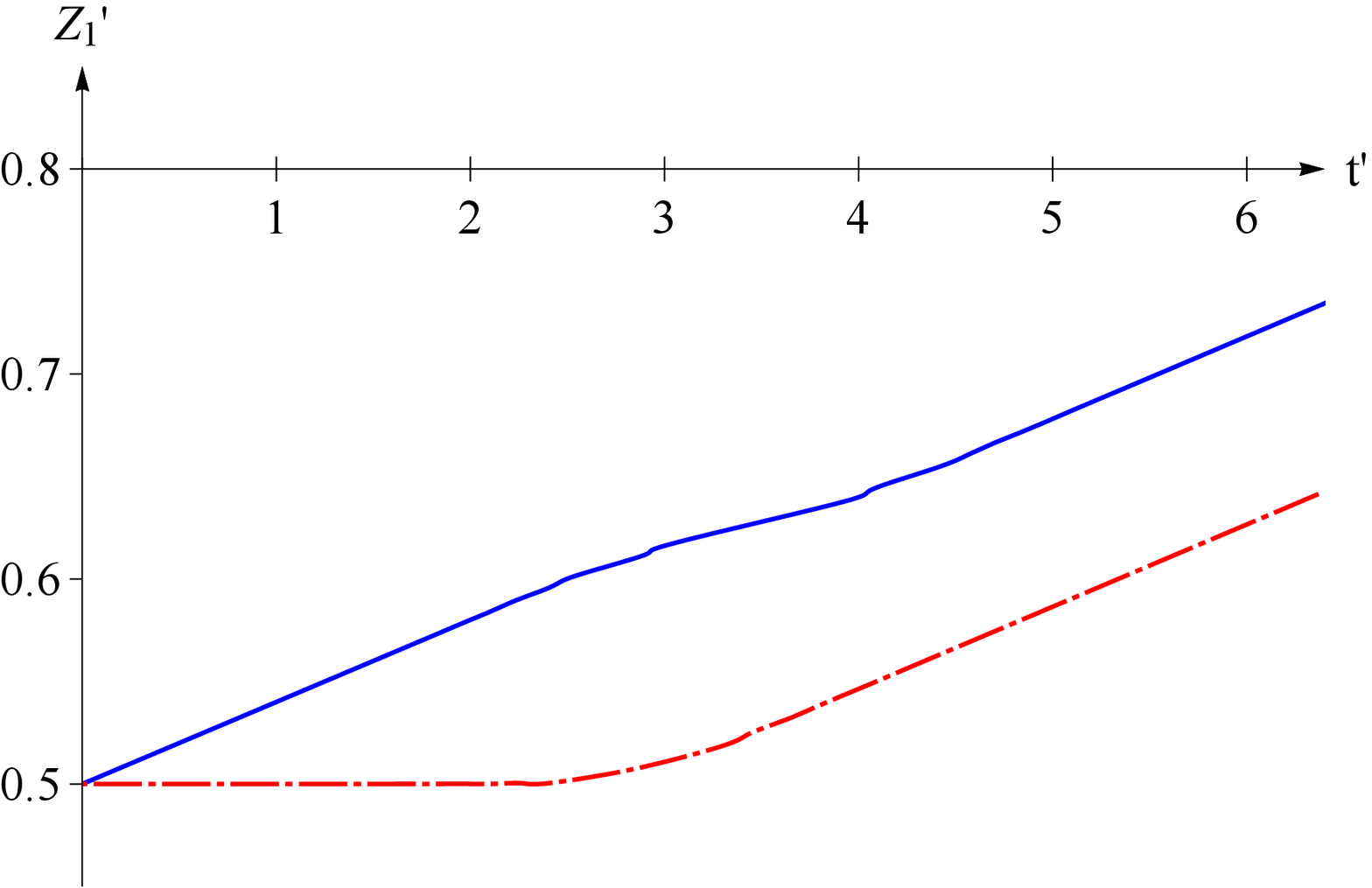}
\end{minipage}\hfill
\begin{minipage}[c]{0.5\textwidth}
\includegraphics[width=5cm]{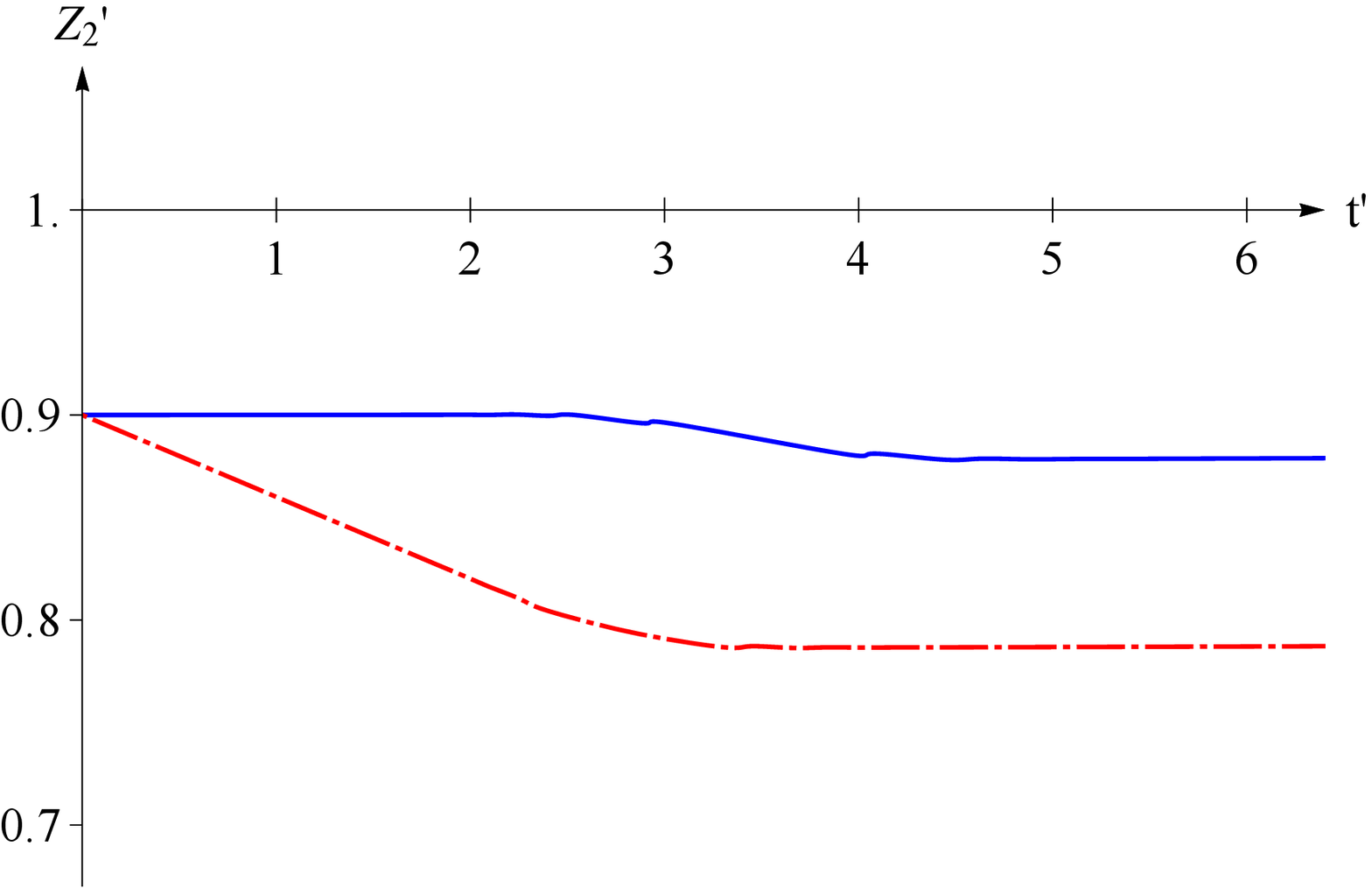}
\end{minipage}
\caption{(color on line) Same figure as Fig. \ref{Fig-7}, but with different initial values for the positions of the pointers ($Z_{1,0}^{\prime} = 0.5 $ and $Z_{2,0}^{\prime}=0.9$) corresponding to a larger average value than in that figure. In this case, after crossing the interference region, the trajectory of the test particle takes a direction that is determined by the initial positions of the pointer particles, revealing a strong effect of the additional variables of the measurement apparatus.}%
\label{Fig-8}%
\end{figure}

The initial values of the Bohmian positions of the pointer particles are of course random. It may happen that they are for instance both positive, and seem to indicate the presence of the test particle in one slit even before the measurement has begun. Figure \ref{Fig-8} shows another example, with the same input parameters as in \ref{Fig-5}, but different initial values of the positions of the pointer particles.  These values are such that, initially, the two pointers have positions corresponding to a detection of the particle in the upper slit.  We then observe the same \textquotedblleft predestination\textquotedblright\ effect as in Figure \ref{Fig-5}: in the interference region, the trajectory of the test particle takes more often a direction that is consistent with this initial value.

This analysis shows that the Bohmian trajectories of the pointer particles always remain consistent with that of the test particle: before the crossing of the interference region, the trajectory of the pointer particles indicates which  slit has been crossed, and after the crossing it indicates in which beam the test particle propagates. Non-local effects are also visible on the pointer trajectories during the crossing time, and constantly reflect the behavior of the test particle. In the words of Ref. \cite{Vaidman-2012}, this phenomenon is a \textquotedblleft dramatic demonstration\textquotedblright\ of the consequences of non-locality on the interpretation of measurements.

\subsection{Single pointer with more particles}

Coming back to the case where the measurement apparatus consists in a single pointer, but assuming that this pointer contains $N$ particles, we have plotted a number of trajectories. The input parameters are the same as in
\S \ \ref{fast-and-slow-pointers}; the only difference is that $N=10$, and
that $10$ initial positions $Z_{n,0}^{\prime}$ of the pointer particles are randomly chosen within their common Gaussian distribution. Since it would be inconvenient to plot $10$ separate trajectories, in the figures we show the trajectory of the averaged variable (also used in the Appendix):
\begin{equation}
\hat{\Sigma}^{\prime} = \frac{1}{\sqrt N} \sum_n  Z_n^{\prime}
\label{defn-sigma-prime}
\end{equation}

\begin{figure}[!b]
\begin{minipage}[c]{0.5\textwidth}
\includegraphics[width=6cm]{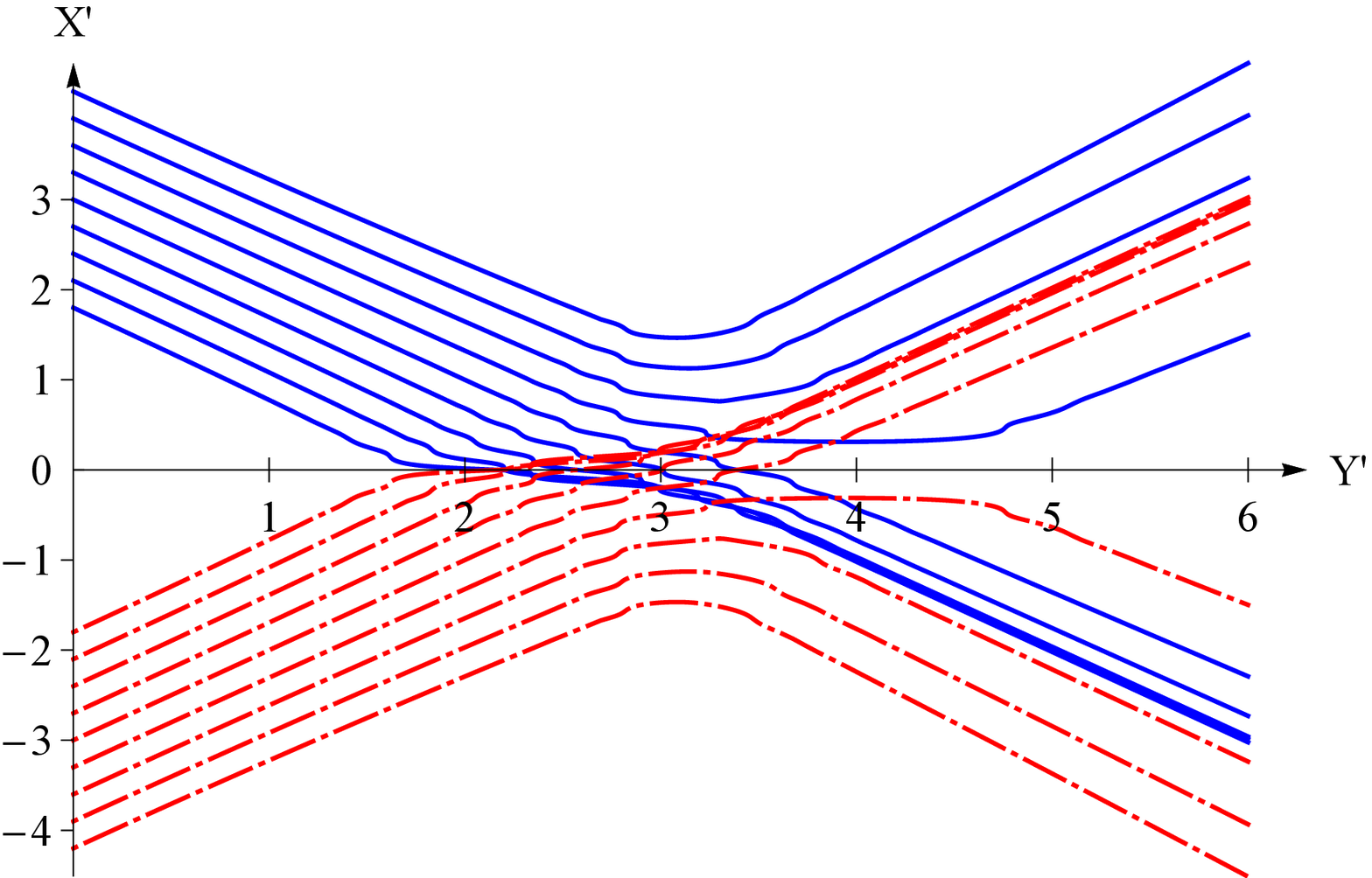}
\end{minipage}\hfill\begin{minipage}[c]{0.5\textwidth}
\includegraphics[width=6cm]{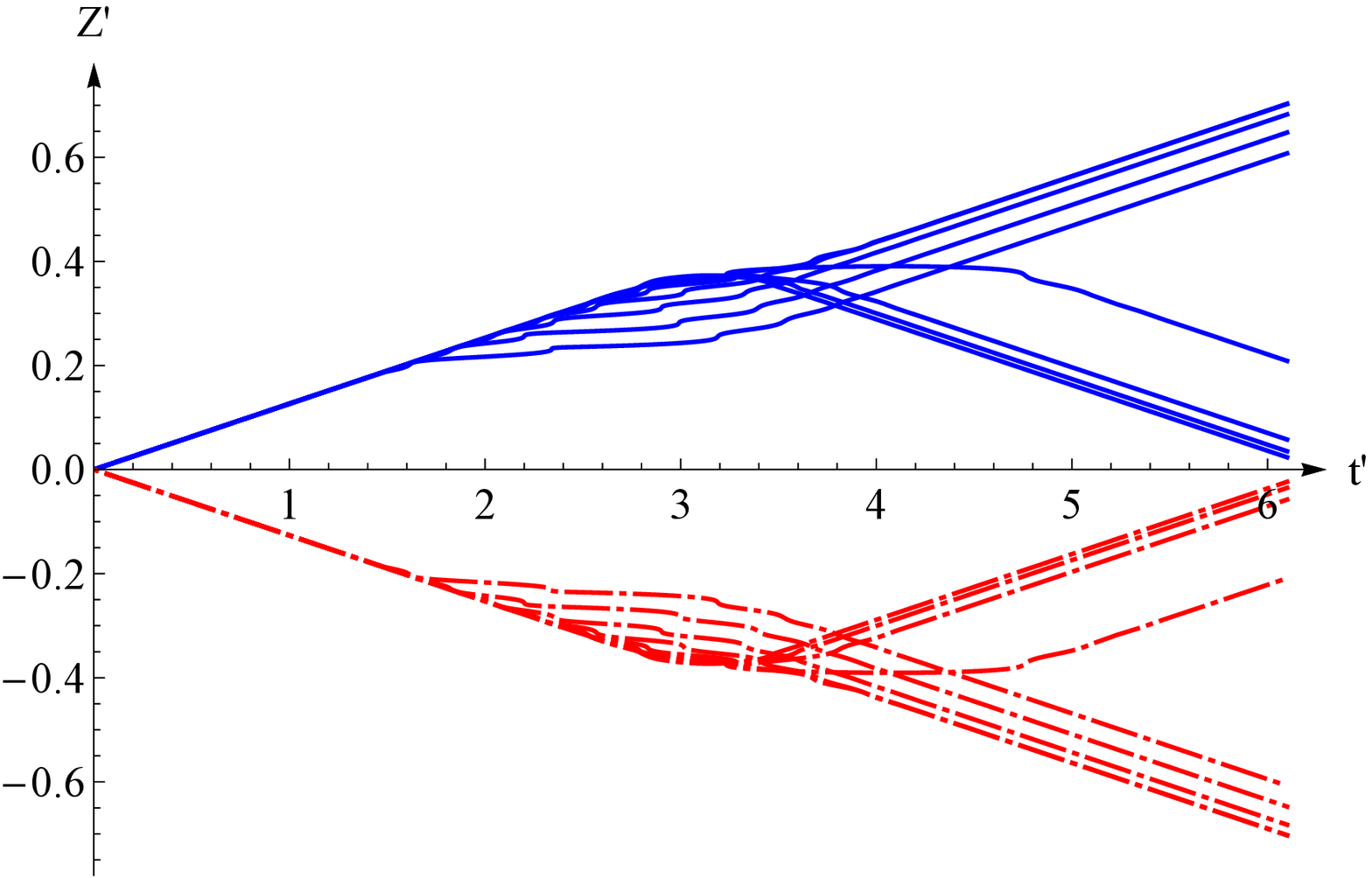}
\end{minipage}
\caption{(color on line) Trajectories obtained with $N=10$ pointer particles. The left part shows the trajectory of the test particle, the right part the trajectory associated with the averaged pointer variable $\hat{\Sigma}^{\prime}$ defined in (\ref{defn-sigma-prime}). This variable is assumed to have an initial value $\hat{\Sigma}^{\prime} (0) =0 $.  Apart from $N$, all parameters are the same as in Figure \ref{Fig-4}. A larger proportion of trajectories than is that figure do not change direction, illustrating the decrease of the proportion of \textquotedblleft surrealistic\textquotedblright\ trajectories when $N$ increases.}%
\label{Fig-9}%
\end{figure}

Figure \ref{Fig-9} shows a case where the initial value of $\hat{\Sigma}^{\prime}$  is zero, a case in which the initial dBB state of
the pointer is neutral (no \textquotedblleft preference\textquotedblright\ for one result or the
other).\ We then observe that the various trajectories of the test particle are
symmetrical with respect to the symmetry plane of the interference
device.\ Some of them still bounce on this plane, as in Figure \ref{Fig-4}, but
a larger proportion of trajectories crosses the interference region: the number of surrealistic
trajectories is therefore reduced when the pointer contains more particles.\

\begin{figure}[!tb]
\begin{minipage}[c]{0.5\textwidth}
\includegraphics[width=6cm]{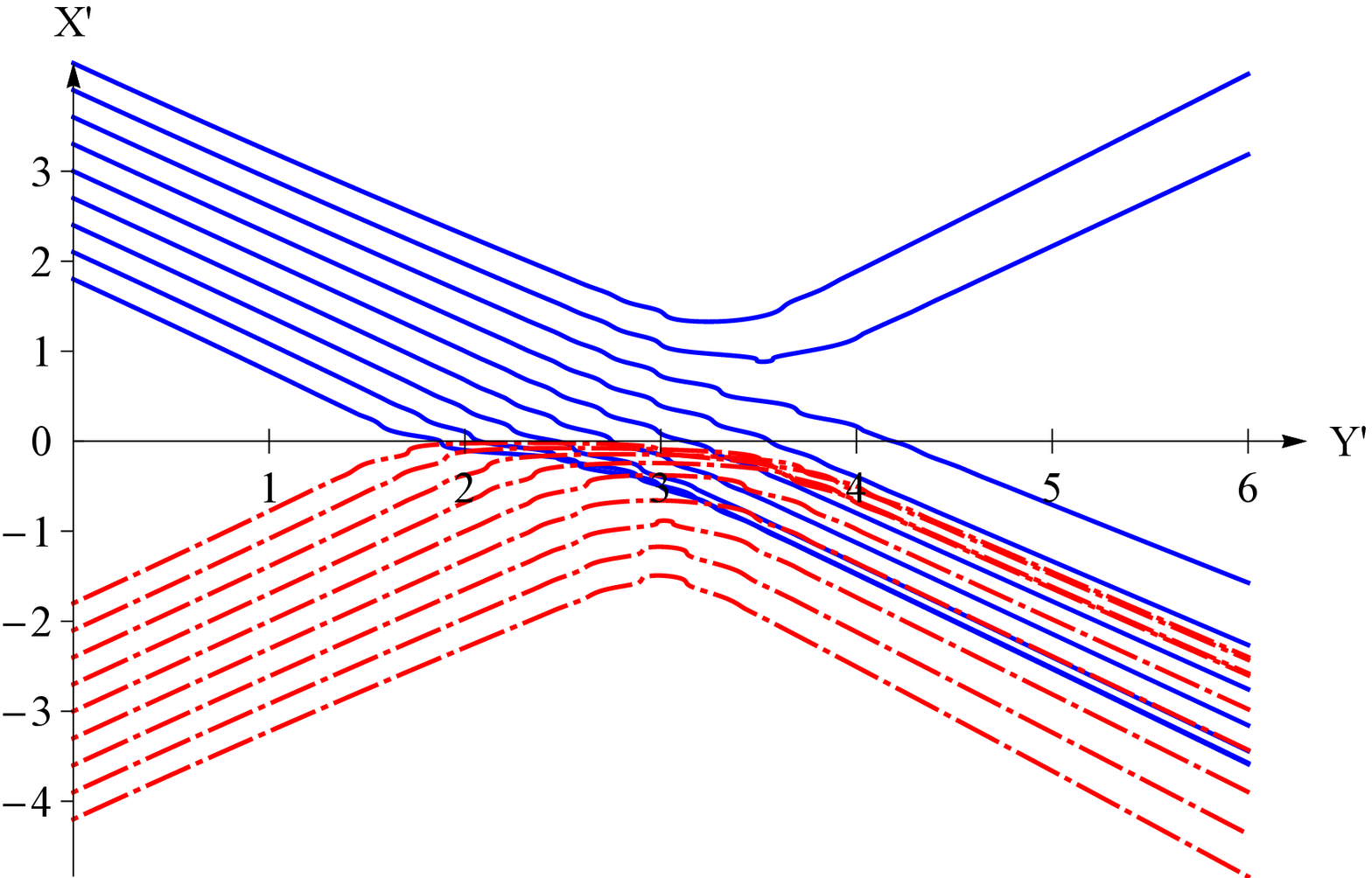}
\end{minipage}\hfill\begin{minipage}[c]{0.5\textwidth}
\includegraphics[width=6cm]{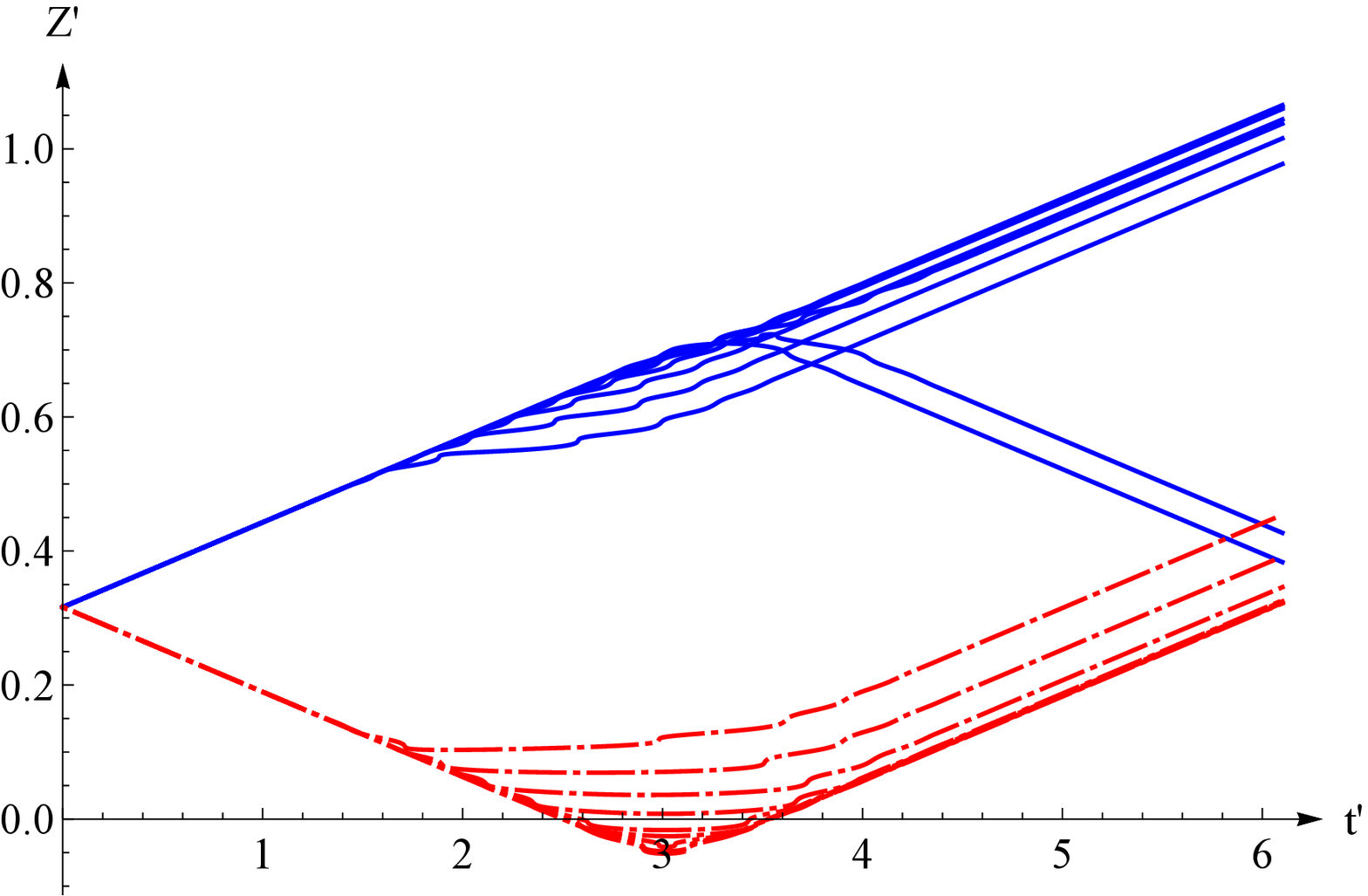}
\end{minipage}
\caption{Trajectories obtained with $N=10$. The initial values of $\hat{\Sigma}^{\prime} (0)$ is now $0.3$, but otherwise all parameters are the same as in Figure \ref{Fig-9}. Because the average value of the positions of the pointer is positive, more trajectories of the test particle seem to originate from the upper than from the lower slit, and go downwards after crossing the interference region.}%
\label{Fig-10}%
\end{figure}

\begin{figure}[!tb]
\begin{minipage}[c]{0.5\textwidth}
\includegraphics[width=6cm]{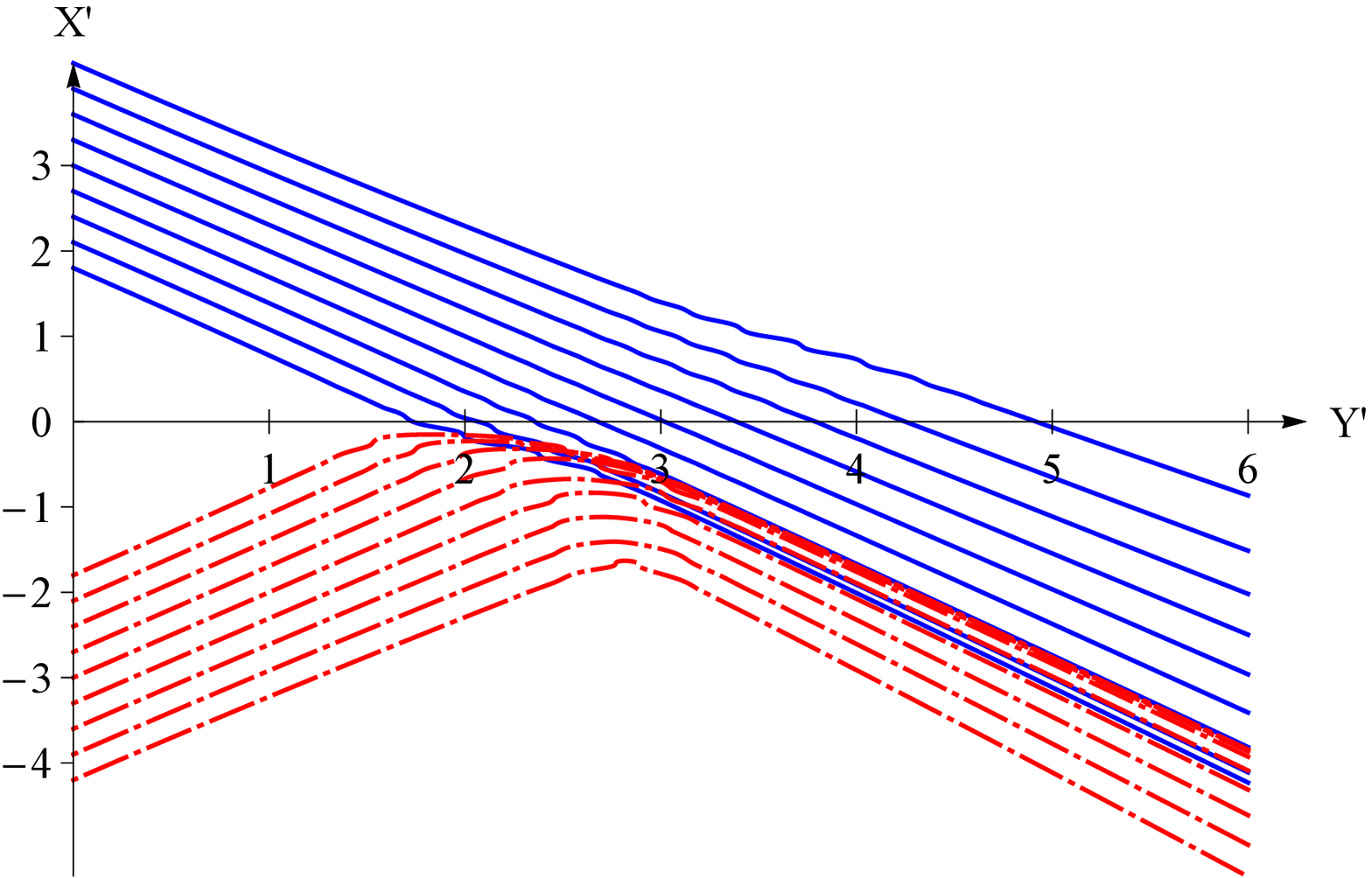}
\end{minipage}\hfill\begin{minipage}[c]{0.5\textwidth}
\includegraphics[width=6cm]{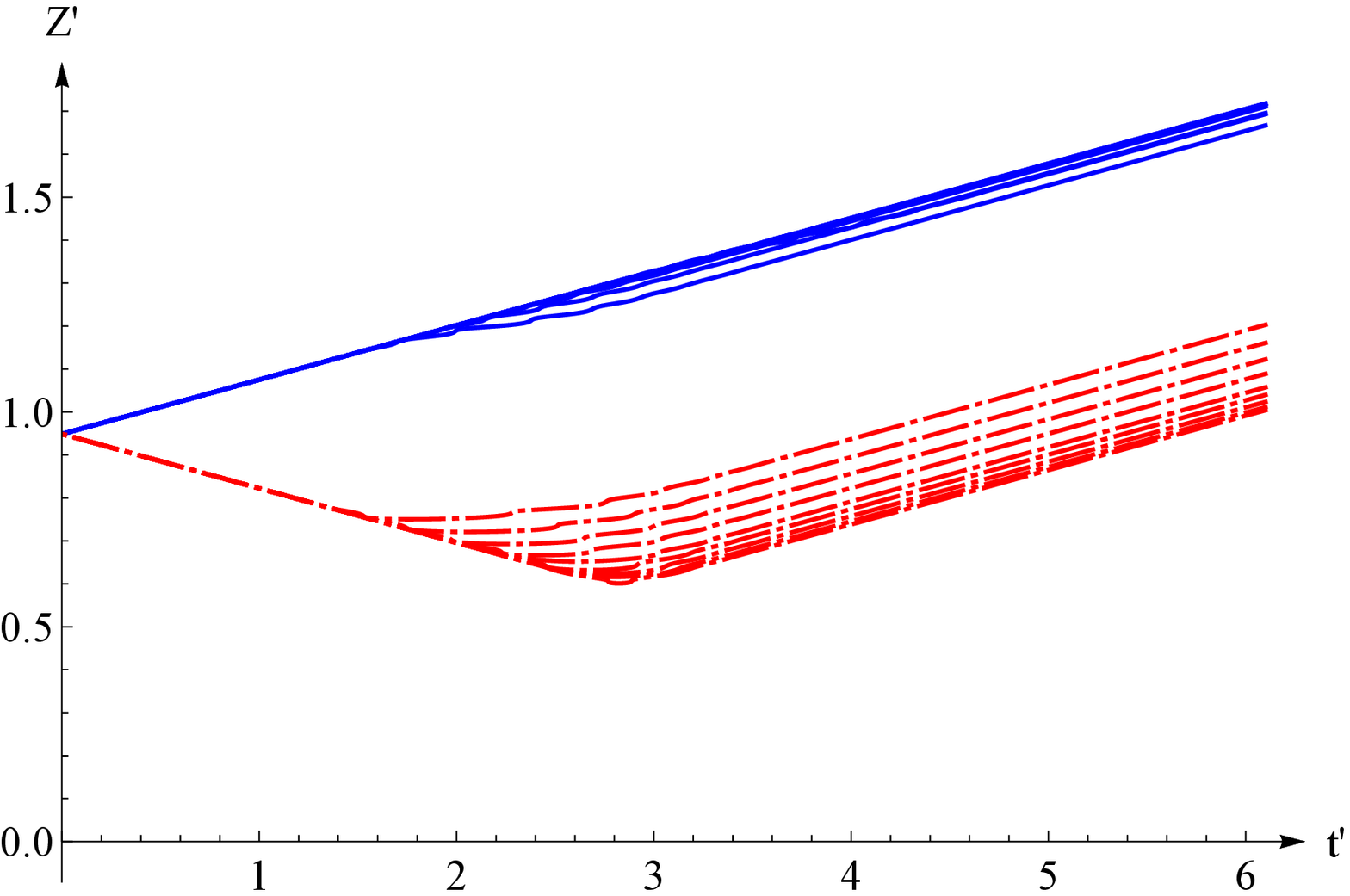}
\end{minipage}
\caption{Same figure as \ref{Fig-10}, but with a still larger initial value $\hat{\Sigma}^{\prime} (0) = 1$. In this case, all trajectories of the test particle after the interference region follow a trajectory that seem to come from the upper slit and goes downwards: the \textquotedblleft result of measurement\textquotedblright\ is determined by the initial values of the additional variables attached to the measurement apparatus (\textquotedblleft predestination effect\textquotedblright ).}%
\label{Fig-11}%
\end{figure}

Figure \ref{Fig-10} shows another case where the initial value of
$\hat{\Sigma}^{\prime}$ is $\hat{\Sigma}^{\prime} (0) =0.3$; this positive value favors one
indication of the pointer and, therefore, a \textquotedblleft result of
measurement\textquotedblright.\ In the interference region, what is observed is that the
trajectories of the test particle tend to deviate
in order to take a direction that seems to agree with this initial average value.
The effect is even more pronounced in Fig. \ref{Fig-11} with a still larger initial value $\hat{\Sigma}^{\prime} (0) =1$; now  almost all trajectories of the test
particle take a final direction that is determined by this initial average.
Interestingly, we have a case where it is not the \textquotedblleft hidden
variable\textquotedblright\ associated with the measured particle that
determines the result of measurement, as often believed in the context of the
dBB theory.\ What matters here is the initial values of the
\textquotedblleft hidden variables\textquotedblright\ of the measurement
apparatus, which determine in advance which path will be taken by the test
particle.\ This sort of \textquotedblleft predestination
effect\textquotedblright\ is also a generalization of the \textquotedblleft
non-local steering of Bohmian trajectories\textquotedblright\ observed with
photons in Ref. \cite{Xiao-et-al-2017}.

\begin{figure}[!tb]
\begin{minipage}[c]{0.5\textwidth}
\includegraphics[width=6cm]{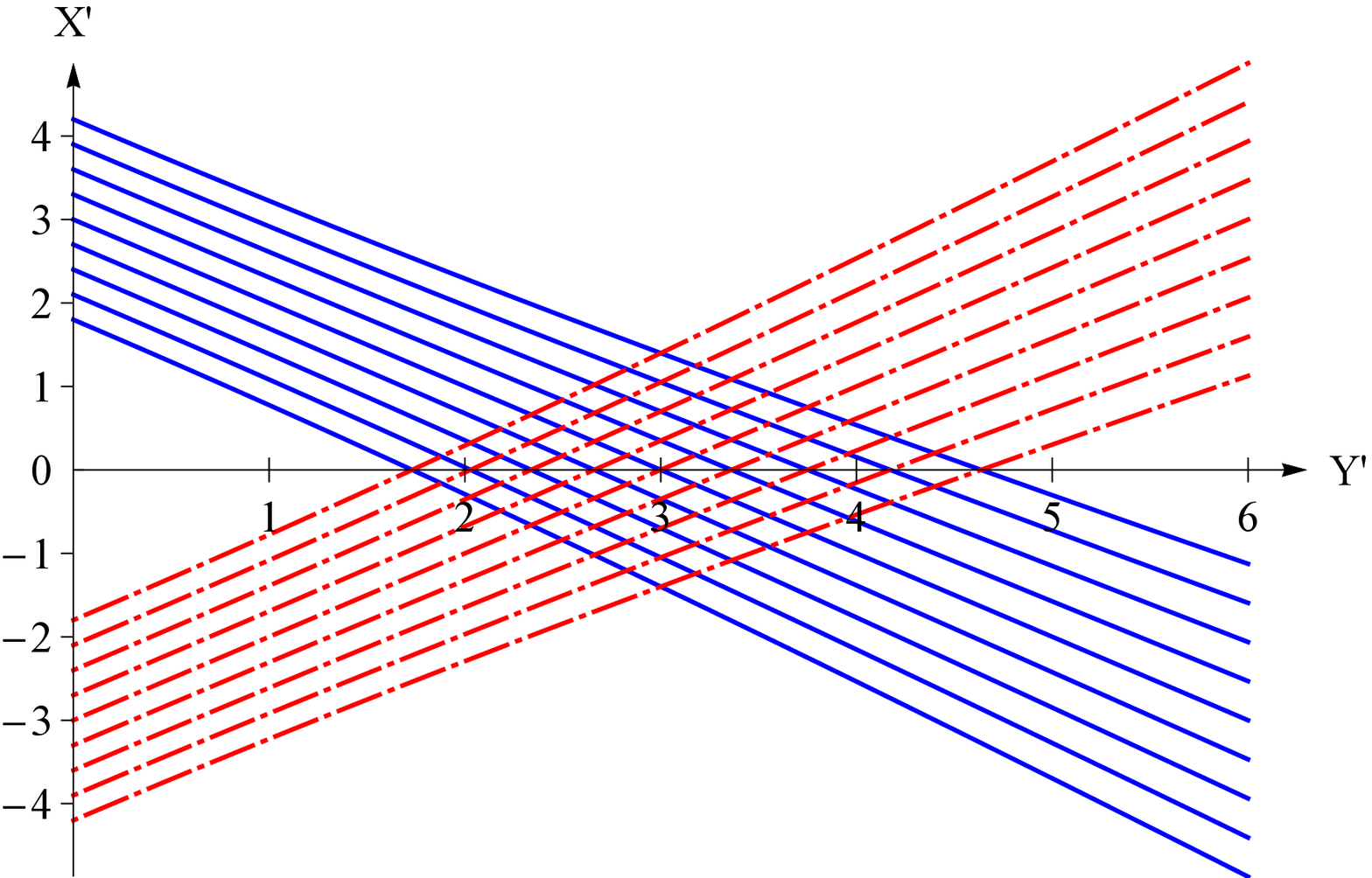}
\end{minipage}\hfill\begin{minipage}[c]{0.5\textwidth}
\includegraphics[width=6cm]{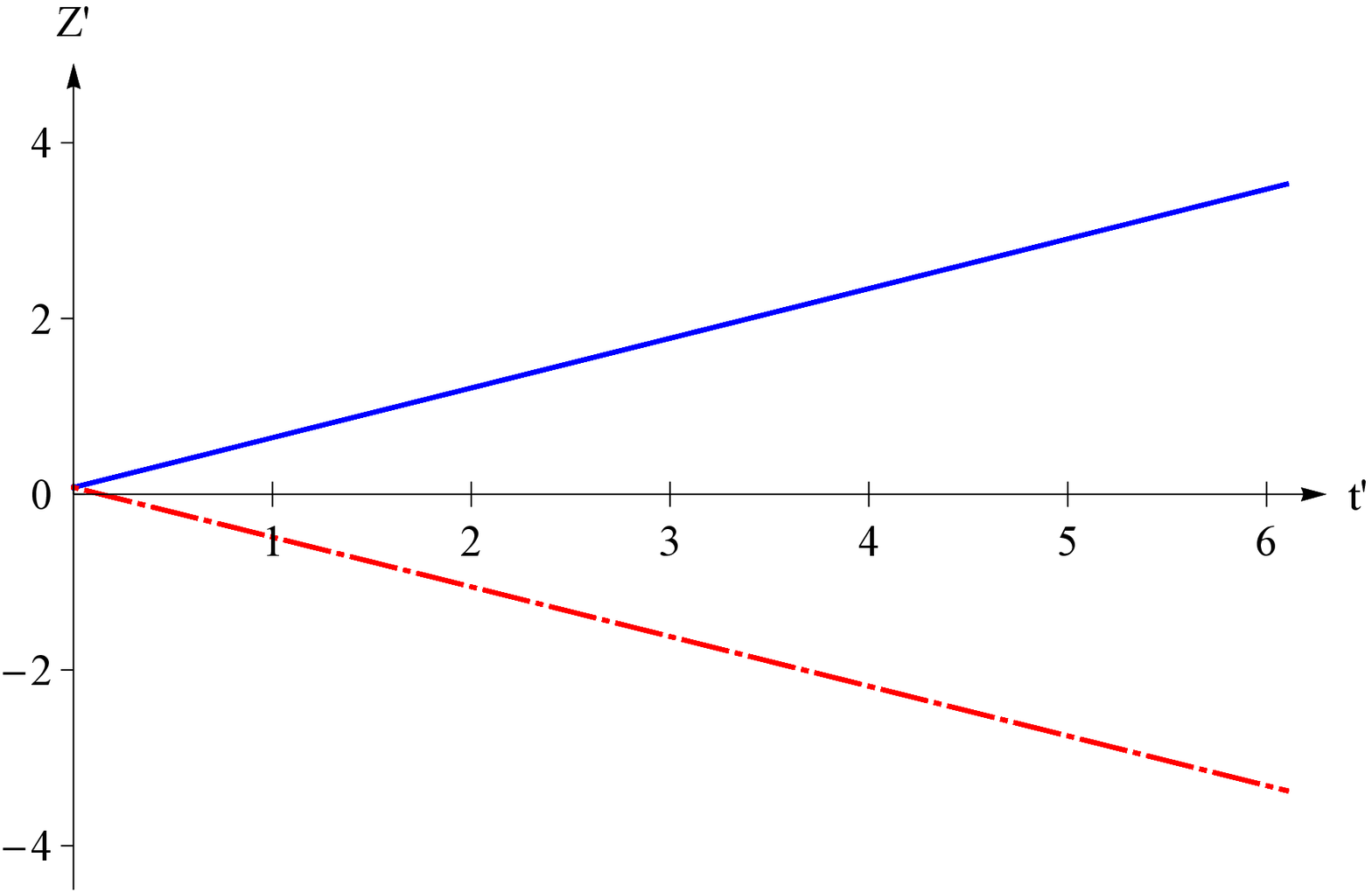}
\end{minipage}
\caption{Trajectories obtained with the same input parameters as those of Fig \ref{Fig-10}, but 200 pointer particles instead of 10. Because the number of pointer particles has increased, the trajectories are now all close to straight lines. The quantum non-local effects have completely disappeared and no Bohmian trajectory can be called \textquotedblleft surrealistic\textquotedblright .}%
\label{Fig-12}%
\end{figure}

We have also performed computations with a larger number of pointer
particles, $N=200$.
Figure \ref{Fig-12} shows the results obtained with input parameters identical to those of Figure
\ref{Fig-11}, but for a pointer composed of $200$ particles.\ Most trajectories of the test particle are now almost straight lines.

The conclusion of this study is that, the larger the number of pointer
particles, the weaker the interference effects of the test particle, and
therefore also the smaller the proportion of apparently surrealistic
trajectories.\ In the limit of very large values of $N$, they completely disappear, as we now show analytically.

\section{Macroscopic pointer, relation to decoherence}
\label{macroscopic-pointer}

We now give a brief analytic argument showing that, when the number of
pointer particles tends to infinity, all trajectories of the test particle
cross the interference region as straight lines, even if the pointer is slow.
This is actually nothing but
the extension of a brief argument given by Bell already in 1980 \cite{Bell-1980}, assuming fast pointers;
it is interesting to note that he had already discussed the essence of the phenomenon 12 years before the notion of surrealistic trajectories was introduced \cite{Englert-1992}.

As we have seen, in order to determine the trajectory of the test
particle in the interference region, the crucial element is the ratio between the two components of
the wave function, evaluated at the Bohmian positions of all particles: if the
Bohmian positions of the pointer are such that only one component remains
active (the other takes negligible values), the Bohmian position of the test
particle follows a straight line, and no surrealistic trajectory occurs.\ We
therefore have to study the $Z_{n}$ dependence of the amplitudes associated
with the two components $\Phi_{\pm}$ given by (\ref{art-1}), which is
contained in the last term of the first line of (\ref{art-100}).

At time $t=0$, when the test particle crosses one of the slits, it is the
Bohmian position of this particle that determines which of the two components
$\Phi_{\pm}$ is effective; the other is inactive, what Bohm calls an
\textquotedblleft empty wave\textquotedblright.\ The Bohmian positions of the
pointer particles have not yet changed, and they give the same values to
the two components of the wave function.\ Nevertheless all these positions
have an initial velocity that depends on the component of the wave function:
$+V$ for the component where the test particles goes through the upper slit,
$-V$ in the other case. The pointer Bohmian positions actually move together with
their wave packet in the effective wave, being insensitive to the empty wave
that would induce an opposite motion.

Consider now a time $\delta t$ that is smaller than the time $t_{\text{cross}%
}$ at which the wave packets of the test particle reaches the interference
region.\ The two wave packets of the pointer are then at a mutual distance
$\delta z=2V\delta t$, so that they may take different values at the Bohmian
positions. Let us first consider just one particle of the pointer labeled $n$.\ The amplitude
of the effective wave function at the position $Z_{n}$ is proportional to :%
\begin{equation}
\left\vert \chi_{a}\left(  Z_{n}\right)  \right\vert \sim\text{e}^{-\left(
Z_{n}\right)  ^{2}/c^{2}} \label{art-103}%
\end{equation}
(we shift the origin for the $Z_n$ coordinate at the center of this wave packet; $c$ is width of the packet, as
defined in \S \ \ref{one-particle-pointer}). The amplitude of the the empty wave is
proportional to:%
\begin{equation}
\left\vert \chi_{b}\left(  Z_{n}\right)  \right\vert \sim
\text{e}^{-\left(  Z_{n}+\delta z\right)  ^{2}/c^{2}} \label{art-104}%
\end{equation}
The ratio between these amplitudes is therefore:%
\begin{equation}
\left\vert \frac{\chi_{b}\left(  Z_{n}\right)  }{\chi_{a}\left(
Z_{n}\right)  }\right\vert =\text{e}^{-2Z_{n}~\delta z/c^{2}}\text{e}%
^{-\left(  \delta z/c\right)  ^{2}} \label{art-105}%
\end{equation}

If one takes into account all particles in the pointer, the ratio $K$
between the amplitudes of the effective and empty waves becomes:%
\begin{align}
K  &  =\exp\left\{  -\sum_{n=1}^{N}\left[  \frac{2Z_{n}\delta z}{c^{2}%
}+\left(  \frac{\delta z}{c}\right)  ^{2}\right]  \right\} \nonumber\\
&  =\exp\left\{  -\frac{N}{c^{2}}\left[  2\left\langle Z\right\rangle \delta
z+\left(  \delta z\right)  ^{2}\right]  \right\}  \label{art-106}%
\end{align}
where $\left\langle Z\right\rangle $ is the average of the Bohmian positions:%
\begin{equation}
\left\langle Z\right\rangle =\frac{1}{N}\sum_{n}Z_{n} \label{art-107}%
\end{equation}
The Bohmian positions take random values; $\left\langle Z\right\rangle $
fluctuates from one realization of the experiment to the next.\ Nevertheless,
when $N$ is large we have:%
\begin{equation}
\left\langle Z\right\rangle \simeq\frac{c}{\sqrt{N}} \label{art-108}%
\end{equation}
so that, if $N>c^{2}/\delta z^{2}$ :%
\begin{equation}
K\simeq\exp\left\{  -N\frac{\left(  \delta z\right)  ^{2}}{c^{2}}\right\}
=\exp\left\{  -4N\frac{\left(  V\delta t\right)  ^{2}}{c^{2}}\right\}
\label{art-109}%
\end{equation}
A high value of $N$ can therefore reduce the values of the empty wave at the
Bohmian positions of the pointer by a large factor: if the pointer contains
$10^{20}$ particles, a factor $10^{20}$ enters in the exponential!

More precisely, the empty wave takes negligible values as soon as:%
\begin{equation}
\delta t\geq\tau=\frac{c}{V\sqrt{N}} \label{art-110}%
\end{equation}
The time during which the $N$ particle wave packets still overlap has significantly decreased: it is
reduced by a factor $\sqrt{N}$, which can be $10^{10}$ or more. With a
macroscopic pointer, in practice this time is always shorter than the
time at which the test particle reaches the interference region.\ Then, even
if the Bohmian position of a single pointer particle can no longer cancel the
empty wave, the cumulative effect of the Bohmian positions will perform this
task. The occurrence of surrealistic trajectories would require an extremely
unlikely distribution of the positions where their average value $\left\langle
Z\right\rangle $ is $\left\langle Z\right\rangle \simeq-\delta z/2$, in other
words an impossible situation when $N$ tends to infinity.

The variables of the pointer play the role of an environment for the test particle, so that the disappearance of curved trajectories is  reminiscent of a decoherence phenomenon. But, in the situations we have studied, decoherence has already fully taken place from the beginning. This is because, in the two components $\Phi_{\pm}$ of the total wave function, the two initial states of the test particle are quasi-orthogonal (because of their spatial separation); this is also the case of the states of the pointer particles (because they do not overlap in momentum space if $\Xi \gg 1$). Since we have chosen
$\Xi=10$ for all figures (except for Fig.~\ref{Fig-2} where entanglement was canceled by setting $\Xi=0$), the off diagonal elements of the reduced density matrix of the test particle have an initial value close to zero. What we have studied is therefore rather
a post-decoherence effect, a change from a situation where the test particle is still influenced
by two effective waves, to another situation where only one wave plays a role. This can also be seen as a change of the conditional quantum potential acting on the test particle: a relaxation from a large value in the interference region to a practically zero value.
The study of this relaxation effect requires, as we have seen, that all degrees
of freedom of the environment, including their Bohmian positions (the pointer positions in our case), should be taken into account. Related effects are discussed in Ref. \cite{Toros-2016},
which show how averages over the Bohmian positions of the environment leads to a reduction of
the \textquotedblleft effective wave functions\textquotedblright .

The reasoning can be generalized to other symmetric (non-Gaussian)
distributions by using a Taylor expansion of its logarithm around the center
($1/c^{2}$ then becomes the second derivative of the distribution at the
origin); higher moments of the $Z$ distribution may then appear, but the
essence of the results remains similar.\ The evolution at long times can also
be studied: in the Appendix, assuming Gaussian
distributions, we show that the scaling of the characteristic evolution time in $1/\sqrt N$ remains valid at any time. Another generalization is to assume that several pointers
detect the passage of the test particle, for instance one pointer per slit as
in \S ~\ref{two-pointers}; this does not change the structure of the
calculations, and the results remain basically the same.

\section{Conclusion}

\label{discussion}

Our conclusion is that, strictly speaking, \textquotedblleft late measurements
of quantum trajectories\textquotedblright\ cannot really \textquotedblleft
fool a detector\textquotedblright\ \cite{Dewdney-et-al-1993}, or at least not fool the physicist who makes careful observations of this detector: no problem, or
contradiction, occurs if the coupled dynamics of the positions is properly understood. It is true that the dynamics of a microscopic pointer may be complex, so that the interpretation of the results of measurements requires a detailed analysis of how the test particle interacts with the measurement apparatus, and of how the pointer moves as a result of this interaction. In particular, since the motion of a pointer may change its direction, a simple extrapolation to the past may lead to incorrect results. Generally speaking, it is known that the mechanism of empty waves and  conditional wave functions  provides a dBB dynamics for the reduction of the wave function; the examples we have studied show that this dynamic may be rather rich.

If the pointer is macroscopic, the final result remains simple: it always behaves as a \textquotedblleft fast pointer\textquotedblright , and neither curved Bohmian
trajectories nor non-local effect take place. As we have
seen, this simple behavior is not predicted if the pointer is treated as a single particle
with a single Bohmian position. It remains of course perfectly legitimate to
introduce a collective variable, for instance the position of the center of
mass of the pointer, and to study its evolution.\ But, when a partial trace
operation over all the variables of the pointer is required, it becomes
indispensable to ascribe a Bohmian position to every particle of the pointer
and to study its contribution.\ It is also
necessary to perform a statistics over the initial values of all Bohmian
positions associated with the pointer; the average value of these positions
and their dispersion plays a significant role.
The morale of the story is that, when a trace operation is necessary in quantum mechanics to evaluate the decoherence induced by the pointer on the test particle, in dBB theory all degrees of freedom of the pointer that are traced out must be taken into account, including every Bohmian position. We mention in passing that this is also important if
one wishes to avoid apparent contradictions concerning the measurement of
correlation functions within dBB theory \cite{Morchio, Neumaier, FL-CNVLMQ}.

If the pointer is microscopic (or mesoscopic) and contains a relatively small
number of particles, quantum non-local effects may indeed take place.\ The variety of possible
phenomena is significantly richer than the simple bounce of trajectories on the symmetry plane generally discussed in the literature. Nevertheless,
no contradiction appears between the whole trajectory of the test particle and
the results of measurements (assumed to be contained in the successive
positions of the pointer particles).\ As soon as the position of the test
particle \textquotedblleft jumps\textquotedblright\ from one wave packet to the other, the positions of the pointer
particles do the same, so that their trajectory constantly reflects that of the test particle.\ In fact, all the information is contained in the
pointer trajectories, ensuring the consistency of the dBB interpretation.   An
interesting feature is that, in some cases, the result of measurement is not
determined by the additional variable attached to the measured test particle,
but by the initial values additional variables of the pointer.\ We then have
an interesting \textquotedblleft predestination effect\textquotedblright\
 where the result is determined by the variables of the measurement apparatus
rather than by that of the measured system.

Any impression of surrealism disappears if one understands that the velocity  of
the particles of the pointer provides information about the velocity of the
position of the test particle at the same time; this is what a detailed
analysis of the coupled dynamics of the particle and the pointer shows. Long
after the test particle crossed the interference region, the motion of the
pointer indicates in which beam the test particle propagates at this time, not which slit it
went through in the past.\ Actually, one can even argue that the trajectories
in question are more real than surreal, since their characteristics (including the changes of directions) should be experimentally accessible by observing the successive positions of the pointer particles with sufficient accuracy.

\newpage

\begin{center}
APPENDIX
\end{center}

The wave functions studied in this article have the general form:%
\begin{equation}
\Psi=R_{1}\text{e}^{iS_{1}}+R_{2}\text{e}^{iS_{2}} \label{52}%
\end{equation}
where $R_{1}$, $R_{2}$, $S_{1}$ and $S_{2}$ are real functions of the position
variables of all particles; the square norms of $R_{1}$ and $R_{2}$ are both
assumed to be $1/2$.

\vspace{3mm}
(i) {\bf General expression of the velocity}
\vspace{1mm}

The probability current associated with a particle of mass $m$\ reads:%
\begin{align}
\mathbf{J} &  =\frac{\hslash}{2im}\left\{  \left[  R_{1}\text{e}^{-iS_{1}%
}+R_{2}\text{e}^{-iS_{2}}\right]  \left[  R_{1}\text{e}^{iS_{1}}\left(
i\bm{\nabla}S_{1}+\frac{\bm{\nabla}R_{1}}{R_{1}}\right)  +R_{2}\text{e}%
^{iS_{2}}\left(  i\bm{\nabla}S_{2}+\frac{\bm{\nabla}R_{2}}{R_{2}}\right)
\right]  \right.  \nonumber\\
&  -\left.  \left[  R_{1}\text{e}^{iS_{1}}+R_{2}\text{e}^{iS_{2}}\right]
\left[  R_{1}\text{e}^{-iS_{1}}\left(  -i\bm{\nabla}S_{1}+\frac
{\bm{\nabla}R_{1}}{R_{1}}\right)  +R_{2}\text{e}^{-iS_{2}}\left(
-i\bm{\nabla}S_{2}+\frac{\bm{\nabla}R_{2}}{R_{2}}\right)  \right]  \right\}
\label{53}%
\end{align}
where the gradients are taken with respect to the coordinates of this
particular particle; for the pointer particles, $m$ is replaced by $M$.\ We then have:%
\begin{align}
\mathbf{J} &  =\frac{\hslash}{m}\left\{  \left(  R_{1}\right)  ^{2}%
\bm{\bm{\nabla}}S_{1}+\left(  R_{2}\right)  ^{2}\bm{\bm{\nabla}}S_{2}\right\}
\nonumber\\
&  +\frac{\hslash}{m}R_{1}R_{2}\left\{  \cos(S_{1}-S_{2})\left[
\bm{\nabla}S_{1}+\bm{\nabla}S_{2}\right]  +\sin(S_{1}-S_{2})\left[
\frac{\bm{\nabla}R_{1}}{R_{1}}-\frac{\bm{\nabla}R_{2}}{R_{2}}\right]
\right\}  \label{55}%
\end{align}
The velocity $\mathbf{V}$ of this particle is then given by:%
\begin{align}
\mathbf{V}  &  =\frac{\hslash}{m}\frac{\left(  R_{1}\right)  ^{2}}{\rho}\bm{\nabla}
S_{1}+\frac{\hslash}{m}\frac{\left(  R_{2}\right)  ^{2}}{\rho}\bm{\nabla}
S_{2}\nonumber\\
&  +\frac{\hslash}{m}\frac{R_{1}R_{2}}{\rho}\left\{  \cos(S_{1}-S_{2})\left[
\bm{\nabla} S_{1}+\bm{\nabla} S_{2}\right]  +\sin(S_{1}-S_{2})\left[  \frac{\bm{\nabla}
R_{1}}{R_{1}}-\frac{\bm{\nabla} R_{2}}{R_{2}}\right]  \right\}  \label{56}%
\end{align}
where $\rho$ is the local density:%
\begin{equation}
\rho= \left(  R_{1}\right)  ^{2}+\left(  R_{2}\right)  ^{2}+2R_{1}R_{2}%
\cos(S_{1}-S_{2}) \label{57}%
\end{equation}
The first line of (\ref{56}) gives the average of the velocities
associated with the two waves, with weights given by their intensities. Ref.
\cite{Vaidman-2012} gives a discussion of the approximation where this term
only is included. But other contributions also arise; they
contain, not only the gradients of the phases of the two waves, but also the
gradients of their amplitudes.\ If we set:%

\begin{equation}
S_{1,2}=\bar{S}\pm\frac{\delta S}{2} \label{55-3}%
\end{equation}
we can rewrite the velocity in the more compact form:%
\begin{equation}
\mathbf{V}=\frac{\hslash}{m}\left\{  \bm{\nabla} \bar{S}+\frac{\left(  R_{1}\right)
^{2}-\left(  R_{2}\right)  ^{2}}{2\rho}\bm{\nabla}\delta S+\frac{R_{1}R_{2}}{\rho
}\sin(\delta S)\left[  \frac{\bm{\nabla} R_{1}}{R_{1}}-\frac{\bm{\nabla} R_{2}}{R_{2}%
}\right]  \right\}  \label{56-bis}%
\end{equation}
\bigskip

The only parts of (\ref{56}) of (\ref{56-bis}) that are particle-dependent are
the gradients, which are taken with respect to the coordinates of the particle
under study, and the value of the mass, $m$ or $M$. Otherwise, $R_1$, $R_2$, $\rho$ and $\delta S$ are functions the Bohmian positions $X$, $Y$ and $Z_n$ defined in the configuration space, which have the same values for all particles.\ We notice that, in
these coefficients, the functions $R_{1}$ and $R_{2}$ appear only though their
ratio $\Omega= R_{1} / R_{2}$.

\vspace{3mm}
(ii) {\bf Gaussian wave functions, trajectory of the center of mass}
\vspace{1mm}

For the calculation that follows, it is convenient to rewrite the wave function (\ref{art-1-2}) in the form:
\begin{align}
\varphi_{\pm}^{x}(x;t)  & \sim\exp\left\{  -\frac{a^{2}}{a^{4}+\frac
{4\hslash^{2}t^{2}}{m^{2}}}\left[  x^{2}+\left(  d-v_{x}t\right)  ^{2}%
\mp2x\left(  d-v_{x}t\right)  \right]  \right\}  \nonumber\\
& \times\exp\left\{  \frac{i}{a^{4}+\frac{4\hslash^{2}t^{2}}{m^{2}}}\left[
\frac{2\hslash t}{m}\left[  x^{2}+\left(  d-v_{x}t\right)  ^{2}\right]
\mp\left[  \frac{mv_{x}x}{\hslash}a^{4}+\frac{4\hslash dxt}{m}\right]
\right]  \right\}  \label{nouveau-phi-x}%
\end{align}
We assume that there is only one pointer, and therefore that the parameter $V$ (or $\Xi$) is the same for all pointer particles. The wave function of each pointer particle has a very similar expression, obtained by substituting $z$ to $x$, $M$ to $m$, $c$ to $a$, $V$ to $-v_x$, and finally cancelling $d$ in (\ref{nouveau-phi-x}).

The Bohmian velocities are obtained from the values of the wave functions at the Bohmian positions $x=X$, $y=Y$ and $z_n=Z_n$. We then obtain:%
\begin{equation}
\Omega=\frac{R_{1}}{R_{2}}=\exp\left\{  \frac{4\left(  d-v_{x}t\right)
X}{a^{2}+4\frac{\hslash^{2}t^{2}}{m^{2}a^{2}}}\right\}  \exp\left\{
\sum_{n=1}^{N}\frac{4Vt~Z_{n}}{c^{2}+4\frac{\hslash^{2}t^{2}}{M^{2}c^{2}}%
}\right\}  \label{58}%
\end{equation}
as well as:%
\begin{equation}
\delta S =S_{1}-S_{2}= - 2 \left[  \frac{mv_{x}X}{\hslash}+\frac{4\hslash dXt}{ma^{4}} \right]
\frac{a^4}{a^{4}+4\frac{\hslash^{2}t^{2}}{m^{2}}} ~  X
+\sum_{n=1}^{N}   \frac{2MV}{\hbar} \frac{c^{4}}{c^{4}+4\frac{\hslash^{2}t^{2}}{M^{2}}} ~ Z_{n}
\label{59}%
\end{equation}
A remarkable property is that, with the Gaussian wave packets we consider,
both functions $\Omega$ and $S_{1}-S_{2}$ depend on the variables $Z_{n}$ only
through their sum $\Sigma$:%
\begin{equation}
\Sigma=\sum_{n=1}^{N}Z_{n} \label{59-2}%
\end{equation}
which is nothing but the position of the center of mass of the pointer multiplied by $N$. Reference \cite{Oriols-Benseny} gives a general discussion of the motion of the center of mass in dBB theory.

For the test particle, the gradients appearing in the velocity are obtained by taking the derivative of the phase and modulus of (\ref{nouveau-phi-x}):%
\begin{equation}
\bm{\nabla}_{x} S_{1,2}=\frac{1}{a^{4}+\frac{4\hslash^{2}t^{2}}{m^{2}}}\left[  4 \frac{\hslash tX}{m} \mp\left(
\frac{mv_{x}}{\hslash}+\frac{4\hslash dt}{m}\right)  \right] \label{60}%
\end{equation}
and:%
\begin{equation}
\frac{\bm{\nabla}_{x} R_{1,2}}{R_{1,2}}= - 2 ~\frac{X \mp ( d - v_{x}t)}{a^{2}%
+4\frac{\hslash^{2}t^{2}}{m^{2} a^{2}}} \label{61}%
\end{equation}
The velocity of the test particle is therefore a function of $X$, of the
time, and of $\Sigma$ only through the $\Sigma$ dependence of $\delta S $ and $\Omega$.

For every pointer particle, the gradients are:%
\begin{equation}
\bm{\nabla}_{z_{n}}S_{1,2}=
\frac{1}{c^{4}+4\frac{\hslash^{2}t^{2}}{M^{2}}}
\left[  4 \frac{\hslash tZ}{M} \mp
\frac{MV}{\hslash}  \right]
\label{62}%
\end{equation}
and:
\begin{equation}
\frac{\bm{\nabla}_{z_{n}}R_{1,2}}{R_{1,2}}= -2 ~\frac{\left[  Z_{n}\mp
Vt\right]  }{c^{2}+4\frac{\hslash^{2}t^{2}}{M^{2} c^{2}}} \label{63}%
\end{equation}
To obtain the time evolution of $\Sigma$, we have to add the velocities of all
positions $Z_{n}$, that is the values of these gradients for all values of
$n$. Relation (\ref{56-bis}) then leads to :%
\begin{align}
\frac{\text{d}}{\text{d}t}\Sigma   =\mathbf{V}_{\Sigma}= &  \frac{c^{4}}{c^{4}+4\frac{\hslash^{2}t^{2}}{M^{2}}} \frac{\hslash}{M}
\left[ \frac{4\hslash t}{M c^{4}} ~ \Sigma  + \frac{\Omega ^2 -1}{1+ \Omega ^2 + 2 \Omega \cos ( \delta S )} ~ \frac{M N V  }{\hslash}   \right.
\nonumber \\
& \hspace{4cm}  \left. + \frac{\Omega }{1+ \Omega ^2 + 2 \Omega \cos \delta S } ~ \sin
(\delta S) ~4  \frac{NVt}{c^{2}}  %
\right]
\label{64}
\end{align}
When the sum of the $Z_{n}$ is replaced by $\Sigma$ in (\ref{58}) and
(\ref{59}), all the individual $Z_{n}$ have disappeared from the equations;
a closed system of equations is obtained for the dynamics of $X$  and
$\Sigma$.

\vspace{3mm}
(iii) {\bf Change of variables, effective velocity}
\vspace{1mm}

In (\ref{64}), the parameters $N$ and $V$ appear only through their product
$NV$, but this is not the case in (\ref{58}) and
(\ref{59}). We therefore change variables and set:%

\begin{equation}
\Sigma=\hat{\Sigma}\sqrt{N} \label{65}%
\end{equation}
to obtain:%
\begin{equation}
\Omega=\frac{R_{1}}{R_{2}}=\exp\left\{  \frac{4\left(  d-v_{x}t\right)X}
{a^{2}+4\frac{\hslash^{2}t^{2}}{m^{2}a^{2}}}\right\}  \exp\left\{
\frac{4Vt\sqrt{N} ~\hat{\Sigma}}{c^{2}+4\frac{\hslash^{2}t^{2}}{M^{2}c^{2}}%
}\right\}  \label{65-bis}%
\end{equation}
as well as:%
\begin{equation}
\delta S = - 2
\frac{a^4}{a^{4}+4\frac{\hslash^{2}t^{2}}{m^{2}}} \left[  \frac{mv_{x} X}{\hslash}+\frac{4\hslash dXt}{ma^{4}} \right]  X
+    \frac{2MV \sqrt N}{\hbar} \frac{c^{4}}{c^{4}+4\frac{\hslash^{2}t^{2}}{M^{2}}} ~ \hat{\Sigma}
\label{66}%
\end{equation}
Similarly, (\ref{64}) becomes:%
\begin{align}
\frac{\text{d}}{\text{d}t}\hat{\Sigma}   = & \frac{c^{4}}{c^{4}+4\frac{\hslash^{2}t^{2}}{M^{2}}}  \frac{\hslash}{M}
\left[ \frac{4\hslash t}{M c^{4}} ~ \hat{\Sigma}  + \frac{\Omega ^2 -1}{1+ \Omega ^2 + 2 \Omega \cos ( \delta S )} ~ \frac{M V \sqrt N  }{\hslash}   \right.
\nonumber \\
& \hspace{4cm}  \left. + \frac{\Omega }{1+ \Omega ^2 + 2 \Omega \cos \delta S } ~ \sin
(\delta S) ~4  \frac{V \sqrt N ~t}{c^{2}}  %
\right]
\label{67}
\end{align}
We now obtain a closed system of evolution equations for $X$ and $\hat{\Sigma}$ where the parameters $N$ and $V$ appear only through the
product $V\sqrt{N}$. We therefore obtain the same evolution as for a single
pointer particle of position $\Sigma/\sqrt{N}$, except that the velocity $V$
is multiplied by $\sqrt{N}$.
If $N$ is of the order of $10^{20}$, the pointer always behaves
as a fast pointer, so that no \textquotedblleft surrealistic\textquotedblright\ trajectory can occur.

\end{document}